\documentclass[sigconf]{acmart}
\usepackage{subfigure}
\usepackage{comment}

\AtBeginDocument{%
  }

\copyrightyear{2025}
\acmYear{2025}
\setcopyright{cc}
\setcctype{by}
\acmConference[IOT 2025]{The 15th International Conference on the Internet of Things}{November 18--21, 2025}{Vienna, Austria}
\acmBooktitle{The 15th International Conference on the Internet of Things (IOT 2025), November 18--21, 2025, Vienna, Austria}
\acmPrice{}
\acmDOI{10.1145/3770501.3770505}
\acmISBN{979-8-4007-1595-2/25/11}

\usepackage{pifont}            
\usepackage{wasysym}
\newcommand{\cmark}{\ding{51}} 
\newcommand{\xmark}{\ding{55}} 
\newcommand*\feature[1]{\ifcase#1 -\or\LEFTcircle\or\CIRCLE\fi}
\usepackage{threeparttable}

\begin{document}

\title{Bridging Technical Capability and User Accessibility: 
Off-grid Civilian Emergency Communication}

\author{Karim Khamaisi}
\affiliation{%
  \institution{University of St. Gallen}
  \city{Dornbirn}
  \country{Austria}}
\email{karim.khamaisi@unisg.ch}

\author{Oliver Kamer}
\affiliation{%
  \institution{University of Zürich}
  \city{Zürich}
  \country{Switzerland}}
\email{oliver@kamer.ch}

\author{Bruno Rodrigues}
\affiliation{%
  \institution{University of St. Gallen}
  \city{Dornbirn}
  \country{Austria}}
\email{bruno.rodrigues@unisg.ch}

\author{Jan von der Assen}
\affiliation{%
  \institution{University of Zürich}
  \city{Zürich}
  \country{Switzerland}}
\email{vonderassen@ifi.uzh.ch}

\author{Burkhard Stiller}
\affiliation{%
  \institution{University of Zürich}
  \city{Zürich}
  \country{Switzerland}}
\email{stiller@ifi.uzh.ch}

\renewcommand{\shortauthors}{Khamaisi et al.}

\begin{abstract}
During large-scale crises disrupting cellular and Internet infrastructure, civilians lack reliable methods for communication, aid coordination, and access to trustworthy information. This paper presents a unified emergency communication system integrating a low-power, long-range network with a crisis-oriented smartphone application, enabling decentralized and off-grid civilian communication. Unlike previous solutions separating physical layer resilience from user layer usability, our design merges these aspects into a cohesive crisis-tailored framework.

The system is evaluated in two dimensions: communication performance and application functionality. Field experiments in urban Zürich demonstrate that the 868 MHz band, using the LongFast configuration, achieves a communication range of up to 1.2 km with 92\% Packet Delivery Ratio, validating network robustness under real-world infrastructure degraded conditions. In parallel, a purpose-built mobile application featuring peer-to-peer messaging, identity verification, and community moderation was evaluated through a requirements-based analysis. 

\end{abstract}




\maketitle

\section{Introduction} \label{chp:introduction}

Mobile applications are utilized for government-to-population crisis communication (\textit{e.g.}, \textit{AlertSwiss} in Switzerland~\cite{alertSwissApp}), relying heavily on centralized networks like Wi-Fi or cellular networks.
However, centralized network infrastructure requires maintenance, planning, and operational management~\cite{manetChallenges2004} and is susceptible to disruptions from natural disasters (earthquakes, hurricanes) and large-scale cyber-attacks, which can destabilize essential services~\cite{manetChallenges2004}.

Infrastructure failures (cell towers, power lines, Internet cables) can isolate populations from emergency services. Off-grid networks enable direct smartphone communication, bypassing centralized infrastructure. Decentralized systems facilitate sharing crucial information—survivor locations, medical assistance, relief updates—where both professional first responders and community volunteers become essential  in managing the incident~\cite{cyrenMain}.

Volunteer-based civil response is necessary in large-scale emergencies for several reasons~\cite{2009manetHelpEmergency}: Emergency response teams are typically overwhelmed and may be partially affected by the disaster itself. Population connectivity remains essential for vital information distribution (\emph{e.g.,} water contamination warnings during blackouts) and for providing information to optimize resource allocation.

Unlike previous work~\cite{lowCostMessages, disasteradioMain, lochaMesh, LOCATE_Main} treating network resilience and user interaction separately, this work integrates physical-layer connectivity and application-layer functionality into a unified system for civilian use during infrastructure failure. The system combines a LoRa-based Mobile Ad-hoc Network (MANET) with a crisis-oriented smartphone application. The following contributions are outlined in this paper:

\begin{itemize} 
    \item Design characteristics of an emergency communication system integrating a MANET with smartphone application for resilient, off-grid civilian use during large-scale infrastructure failures.
    \item Real-world field experiments in urban Zürich with up to 10 LoRa nodes, evaluating performance under disaster-like conditions and measuring Packet Delivery Ratio (PDR), coverage range, and signal characteristics.
    \item Functional evaluation of the mobile application examining fulfillment of operational requirements critical for effective civilian emergency usage.
\end{itemize}

Real-world field experiments demonstrate that 868 MHz \textit{LongFast} channel achieves optimal performance with 1.2 km range and 92\% PDR, confirming suitability for dense urban disaster conditions. The mobile application integrating identity verification, community moderation, and resilient messaging achieved a System Usability Scale (SUS) score of 74/100, indicating strong acceptance for untrained civilian responders.

The remainder of this paper is structured as follows: Section~\ref{sec:relatedwork} discusses related work and fundamental concepts.
While Section~\ref{sec:design} introduces the evaluation methodology, metrics, and scenarios, Section~\ref{sec:results} presents the qualitative and quantitative evaluation. Conclusions and future work are outlined in Section~\ref{sec:conclusion}.

\section{Related Work} \label{sec:relatedwork}

\textbf{Radio.} Although LoRa is excellent in line-of-sight (LOS) rural environments~\cite{lowCostMessages, sharma2023lora}, its application in urban emergency communications remains underexplored. Further, in a disaster scenario, both the mobility of nodes and a user-friendly mobile application are crucial. LOCATE, for example, uses LoRa-to-Bluetooth radios with delay-aware routing but lacks non-expert usability evaluation~\cite{LOCATE_Main}. Disasteradio and Locha Mesh target solar-powered rooftops and censorship resistance respectively~\cite{disasteradioMain,lochaMesh}, assuming fixed nodes with minimal mobile tooling for non-experts. In contrast, Meshtastic~\cite{meshtasticFirmware} offers cross-platform firmware and an active community, making it a practical base for mobile, volunteer-driven networks—the focus of our paper.

Non-LoRa alternatives for emergency systems include Wi-Fi MANET prototypes~\cite{MANETCommunication}, and Automatic Packet Reporting System (APRS) amateur-radio links \cite{ARPSPoC}. Wi-Fi MANETs suffer from short range and high power draw, while APRS requires licensed operators, hindering spontaneous civilian participation.

Classic MANET efforts explore heterogeneous backbones for disasters. Luglio \textit{et al.} combine a terrestrial MANET with satellite links but require specialized gateways and terminals~\cite{luglio_interworking_2007}. Jang \textit{et al.} build a Wi-Fi mesh (“P2Pnet”) from rescue-volunteer laptops, assuming mains-powered notebooks and short-range radios~\cite{jang_rescue_2009}. Verma and Chauhan add an energy-aware overlay atop residual cellular coverage~\cite{verma_manet_2015}, while Dey \textit{et al.} reduce broadcast storms with a counter-based scheme evaluated only in simulation~\cite{dey_1_2019}. None validate low-power long-range radio with modern mobile interfaces.

\textbf{Mobile application.} Several commercial apps provide infrastructure free messaging. AirChat uses Bluetooth only \cite{difranco_airchat_2023}. FireChat combined Bluetooth and Wi-Fi mesh and was widely adopted during the 2014 Hong Kong and 2019 Iraq protests before its removal from app stores \cite{hern_firechat_2014,shadbolt_firechat_2014}. Bridgefy offers similar offline messaging, but independent audits show vulnerabilities to tracking, forgery, and plaintext recovery \cite{bridgefy_inc_bridgefy_2023}. HypeSDK (HypeLabs) exposes a cross-radio API for developers \cite{hypelabs_inc_hypelabs_2020}. Most of these solutions rely on Apple’s Multipeer Connectivity framework \cite{mattt_multipeer_2013}, which restricts them to iOS only solutions and to short-range radios (<100 m).  Successful MANETs like FireChat \cite{hern_firechat_2014} and Bridgefy \cite{bridgefy_inc_bridgefy_2023} remain closed-source, while many academic proposals \cite{verma_manet_2015, jang_rescue_2009, luglio_interworking_2007} lack published implementations.

\tablename~ \ref{tab:related} shows off-grid communication project landscapes. Long-range systems (\textit{e.g.,} LoRa or satellite) rarely support mobile, non-expert operation, while usability-oriented apps are confined to short-range Bluetooth/Wi-Fi with security or platform constraints. Our proposal addresses this gap by pairing Meshtastic-based LoRa networking with a crisis-oriented smartphone application, validated through field trials and human-centered testing in dense urban disaster scenarios.

\begin{table}[t]
\centering

\caption{Representative off-grid communication projects and their key characteristics}
\resizebox{\columnwidth}{!}{%
    \begin{tabular}{l l c c l}
    \toprule
    \textbf{System/Project} & \textbf{Radio/Backbone} & \textbf{Range$^\dagger$} & \textbf{Mobile Solution} & \textbf{Major limitation} \\
    \midrule
    LOCATE \cite{LOCATE_Main}          & LoRa $\rightarrow$ Bluetooth & $\sim$1\,km   & \cmark & No non-expert evaluation         \\
    Disasteradio \cite{disasteradioMain} & LoRa (solar nodes)           & $\sim$1–2\,km & \xmark & Fixed rooftop deployment         \\
    Locha Mesh \cite{lochaMesh}        & LoRa (censorship focus)      & $\sim$1–2\,km & \xmark & Minimal mobile tooling           \\
    Meshtastic \cite{meshtasticFirmware} & LoRa (open firmware)         & $\sim$1–4\,km & \cmark & Usability not formally tested    \\ \hline
    Wi-Fi MANET \cite{MANETCommunication} & Wi-Fi ad hoc                 & $<$0.1\,km    & \cmark & Short range, high power          \\
    APRS \cite{ARPSPoC}                & VHF/UHF amateur              & $>$10\,km     & \cmark & Requires radio licence           \\ \hline
    Luglio \textit{et al.} \cite{luglio_interworking_2007} & MANET + satellite            & multi-km      & \cmark & Needs satellite terminals        \\
    Jang \textit{et al}. \cite{jang_rescue_2009}  & Wi-Fi laptop mesh            & $<$0.1\,km    & \cmark & Mains-powered notebooks          \\
    Verma \& Chauhan \cite{verma_manet_2015} & Hybrid cellular / MANET      & multi-km      & \cmark & Relies on residual cellular      \\
    Dey \textit{et al.} \cite{dey_1_2019}        & MANET (simulation)           & n/a           & \cmark & Simulation only                  \\ \hline
    AirChat \cite{difranco_airchat_2023} & Bluetooth                    & $<$0.05\,km   & \cmark & Very short range                 \\
    FireChat \cite{hern_firechat_2014}   & Bluetooth+Wi-Fi mesh         & $<$0.1\,km    & \cmark & Discontinued, short range        \\
    Bridgefy \cite{paterson_mesh_2021}   & Bluetooth mesh               & $<$0.1\,km    & \cmark & Security vulnerabilities         \\
    HypeSDK \cite{hypelabs_inc_hypelabs_2020} & Bluetooth/Wi-Fi API          & $<$0.1\,km    & \cmark & iOS-centric, short range         \\
    \bottomrule
    \end{tabular}}
\begin{flushleft}
\footnotesize $^\dagger$Indicative maximum one-hop range reported in literature or project documentation.
\end{flushleft}
\label{tab:related}
\end{table}


\begin{figure*}[h]
        \centering
        \subfigure[]{\includegraphics[trim= 25 20 25 10,clip, width=0.195\textwidth]{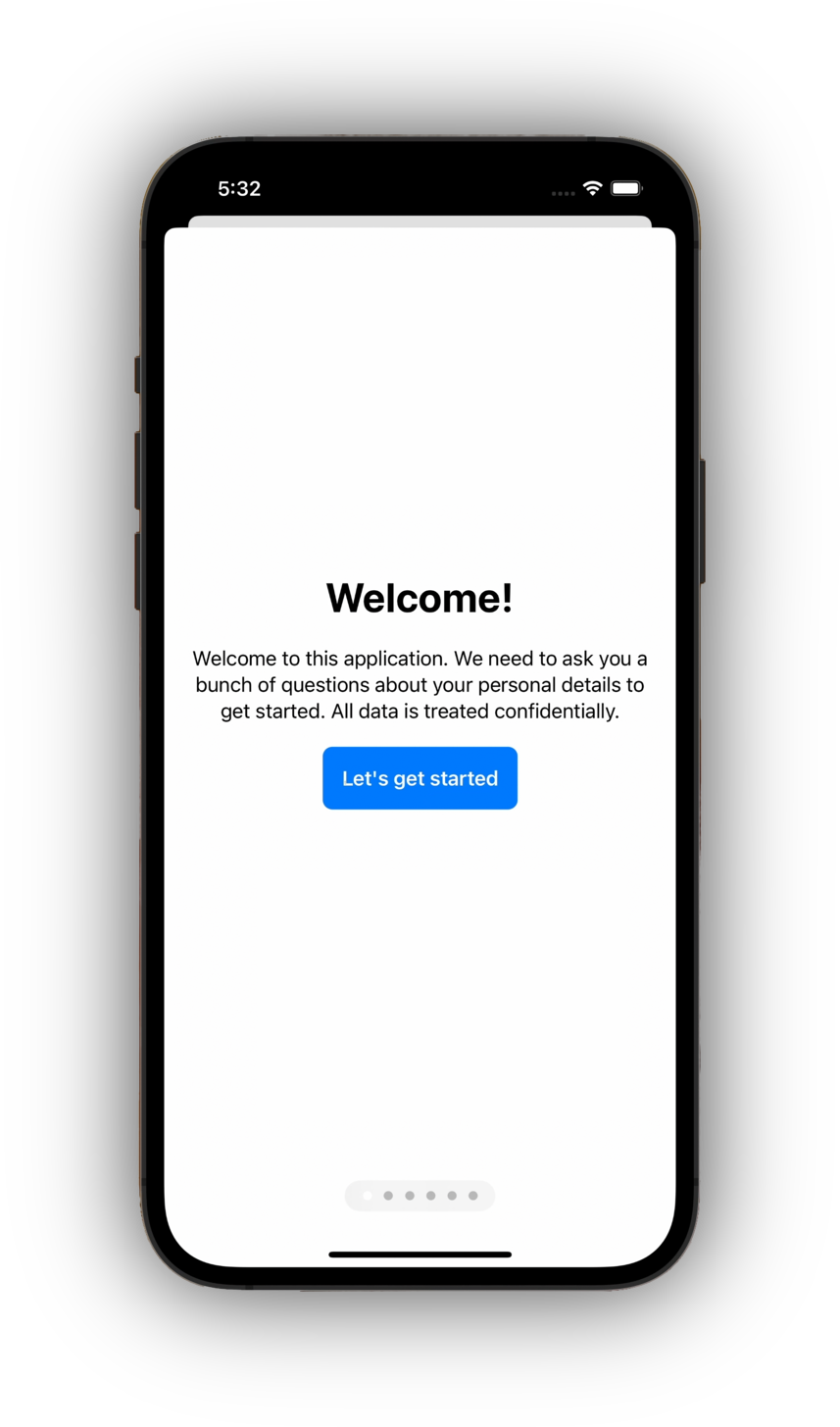}} 
        \subfigure[]{\includegraphics[trim= 25 20 25 10,clip,width=0.195\textwidth]{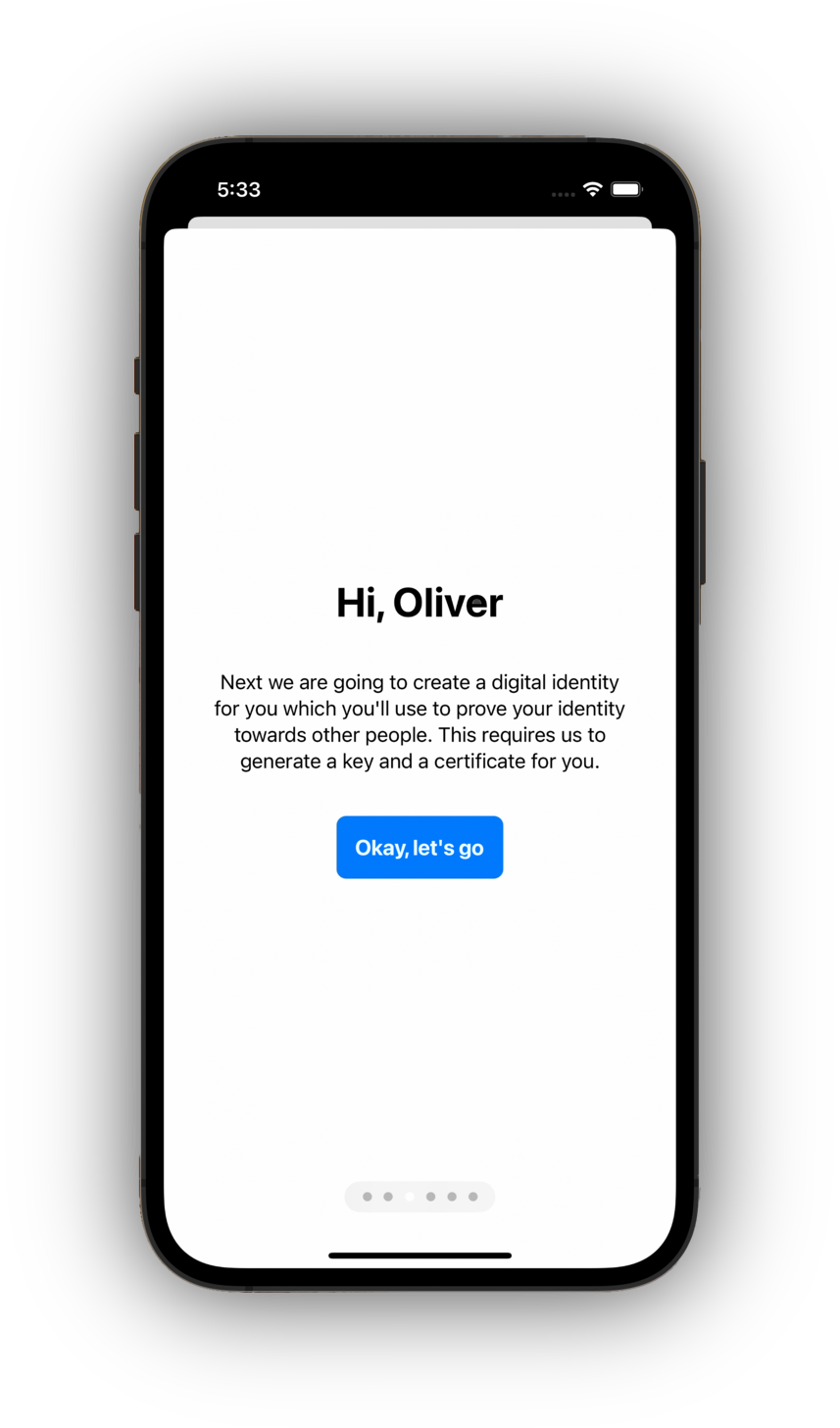}}
        \subfigure[]{\includegraphics[trim= 25 20 25 10,clip,width=0.195\textwidth]{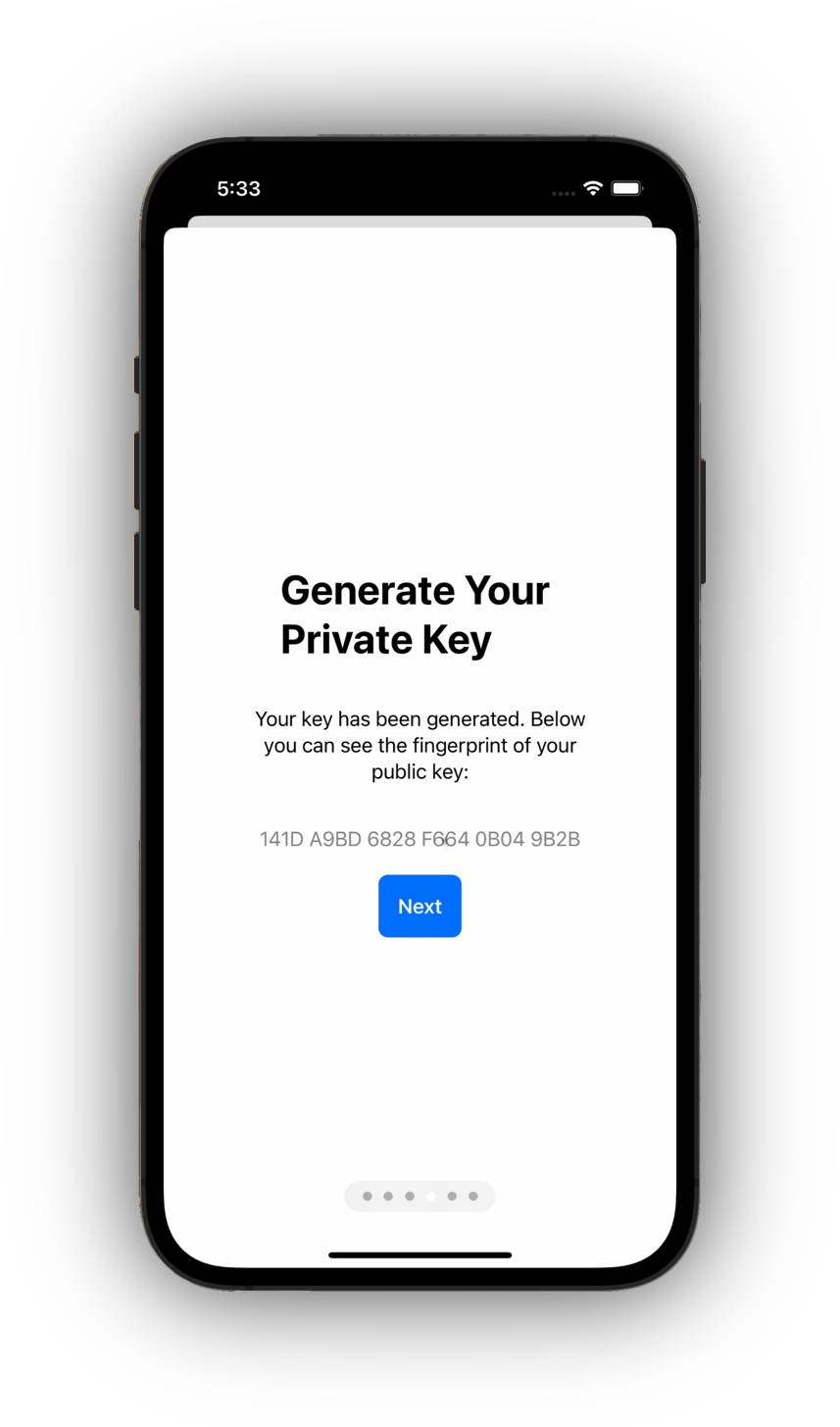}}
        \subfigure[]{\includegraphics[trim= 25 20 25 10,clip,width=0.195\textwidth]{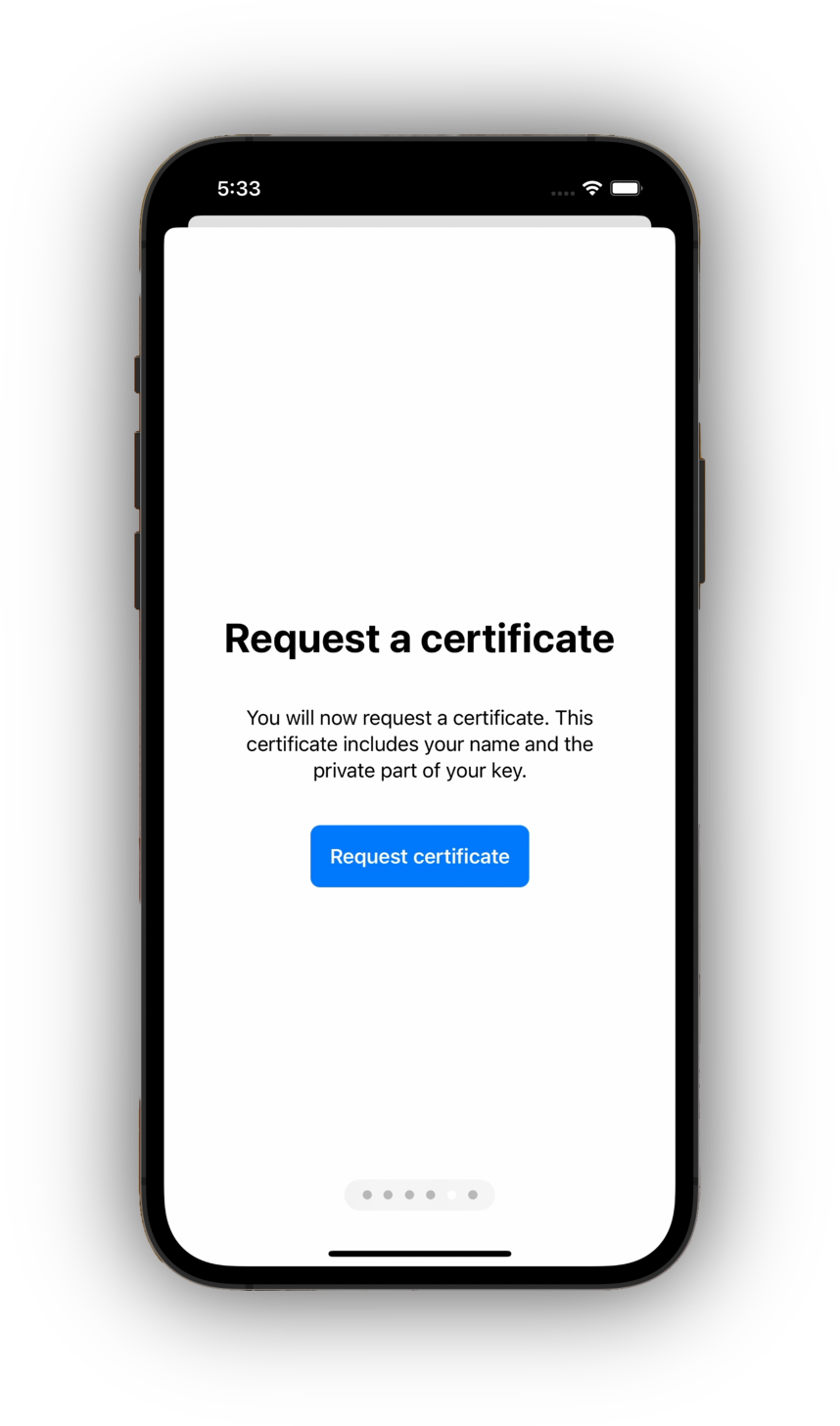}}
        \subfigure[]{\includegraphics[trim= 25 20 25 10,clip,width=0.195\textwidth]{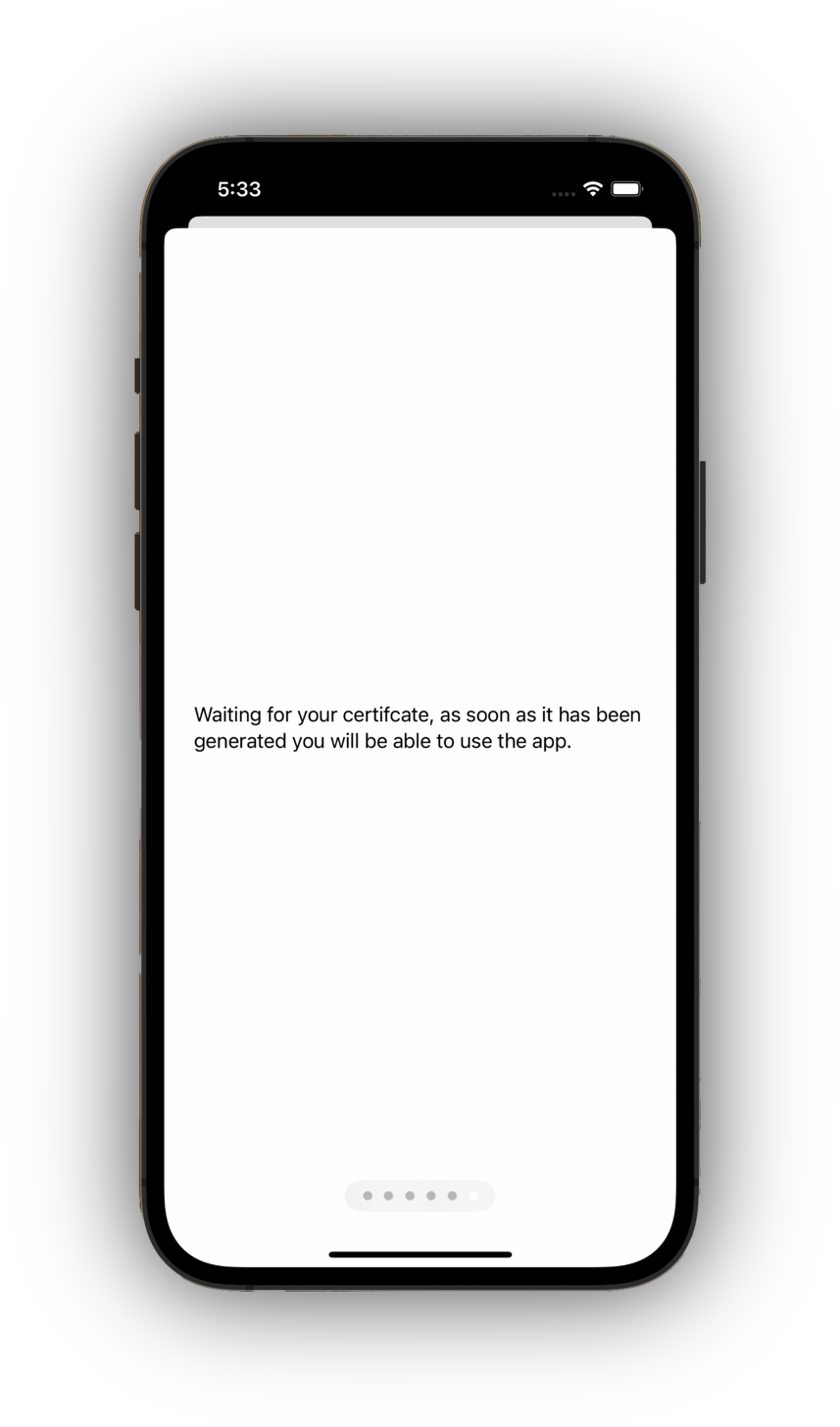}}
        \caption{Identity service during the signup process-flow: (a) Welcome Screen, (b) Digital Identity, (c) Private Key, (d) Certificate Request, (e) Confirmation}
        \label{fig:signup_process}
\end{figure*}

\section{Design} 
\label{sec:design}

\begin{figure}[b]
  \centering
  \resizebox{\columnwidth}{!}{
      \includegraphics[trim= 20 4 20 4,clip]{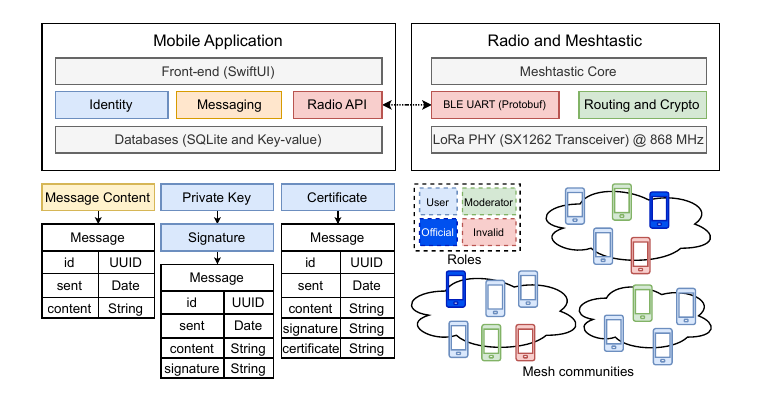}}
  \caption{Component-view: mobile app and radio. Message object (bottom left) and mesh network including user roles (bottom right)}
  \label{fig:architecture}
\end{figure}

The design comprises two major components—mobile application and radio—that create the mesh network (cf. Figure \ref{fig:architecture}). Seven requirements drive the design are presented in \tablename{}~\ref{tab:requirements}: R1-R3 target physical constraints of large-scale emergencies, R4-R5 address human aspects (cross-platform usability, verifiable identities preventing misinformation), and R6-R7 handle temporal emergency dynamics (cellular fallback, easy bootstrapping).
Figure \ref{fig:architecture} presents main design components, message structure, and user roles. The \textbf{identity} service manages certificate-based authentication, enabling message signing and community-moderator roles. The \textbf{messaging} service delivers location-scoped and peer-to-peer chat through the \textit{BLE link} protobuf API~\cite{meshtastic_RadioSettings}. The identity service binds users to verifiable certificates from government-ID checks, while the messaging service delivers signed, tamper-proof chat for specific locations (\textit{e.g.,} postal code "8050").

\begin{table}[h!]
\centering
\caption{Requirements for Emergency Communication}
\label{tab:reqs}
\resizebox{\columnwidth}{!}{%
\begin{tabular}{l p{3cm} p{3.5cm} p{4.1cm} p{4.1cm}}
\toprule
\textbf{ID} & \textbf{Requirement} & \textbf{Emergency need} & \textbf{Gap in prior work$^\ddagger$} & \textbf{Our approach} \\
\midrule
R1 & Kilometre-scale, licence-free radio & Bridge blocked areas when networks fail & P2P apps $<\!100$ m; Wi-Fi MANETs power-hungry \cite{hern_firechat_2014,MANETCommunication} & LoRa 868 MHz LongFast (1.2 km, 92 \% PDR) \\
R2 & Phone-enabled use, no radio skills & Civilians lack RF expertise/licences & APRS needs licences \cite{ARPSPoC}; LOCATE skips user tests \cite{LOCATE_Main} & BLE bridge + Meshtastic \\
R3 & 24 h node on power-bank battery & Grid down; volunteers carry USB packs & Wi-Fi laptop meshes drain power \cite{jang_rescue_2009} & ESP32-S3 + SX1262, $<0.4$ W avg. \\
R4 & Cross-platform app, \emph{good usability} & Mixed device fleet; ease adoption & Few user studies \& iOS-only tools \cite{mattt_multipeer_2013} & Requirement-based analysis \\
R5 & Verified identity + moderation & Limits misinformation, abuse & Bridgefy spoofing flaws \cite{paterson_mesh_2021} & PKI, role-based moderation \\
R6 & Optional cellular fallback & Use LTE/5G to offload LoRa & LoRa projects lack switchover & Auto-select infrastructure in app \\
R7 & Bootstrap in under 5 min & First hour is critical & Satellite hybrids need special kit \cite{luglio_interworking_2007} & Pre-flashed radios, QR key exchange \\
\bottomrule
\end{tabular}}
\begin{flushleft}
\footnotesize $^\ddagger$Citations indicate where the requirement is only partially met in earlier systems.
\end{flushleft}
\label{tab:requirements}
\end{table}

\begin{figure*}[ht]
        \centering
        \subfigure[]{\includegraphics[width=0.185\textwidth]{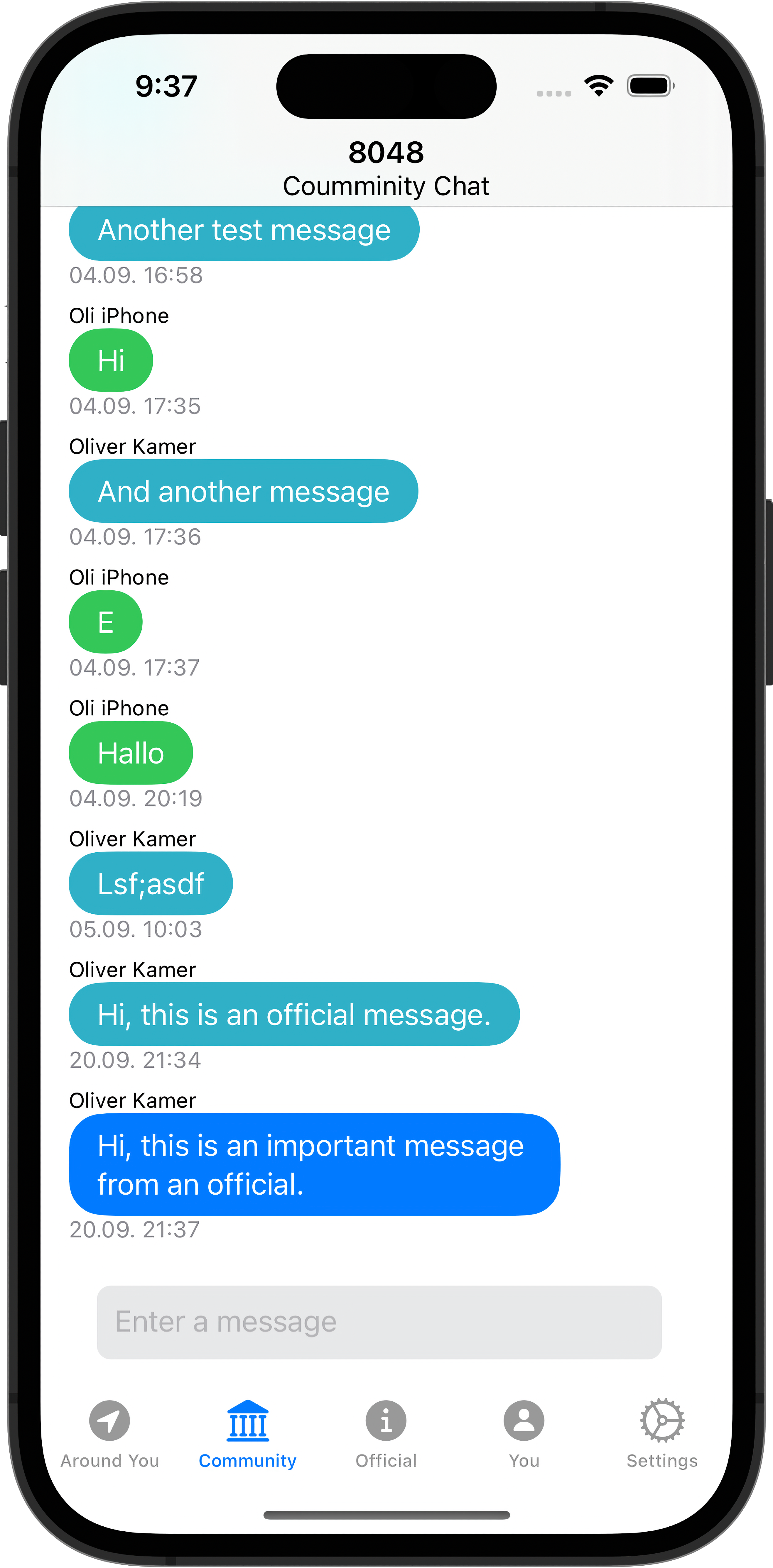}} 
        \subfigure[]{\includegraphics[width=0.185\textwidth]{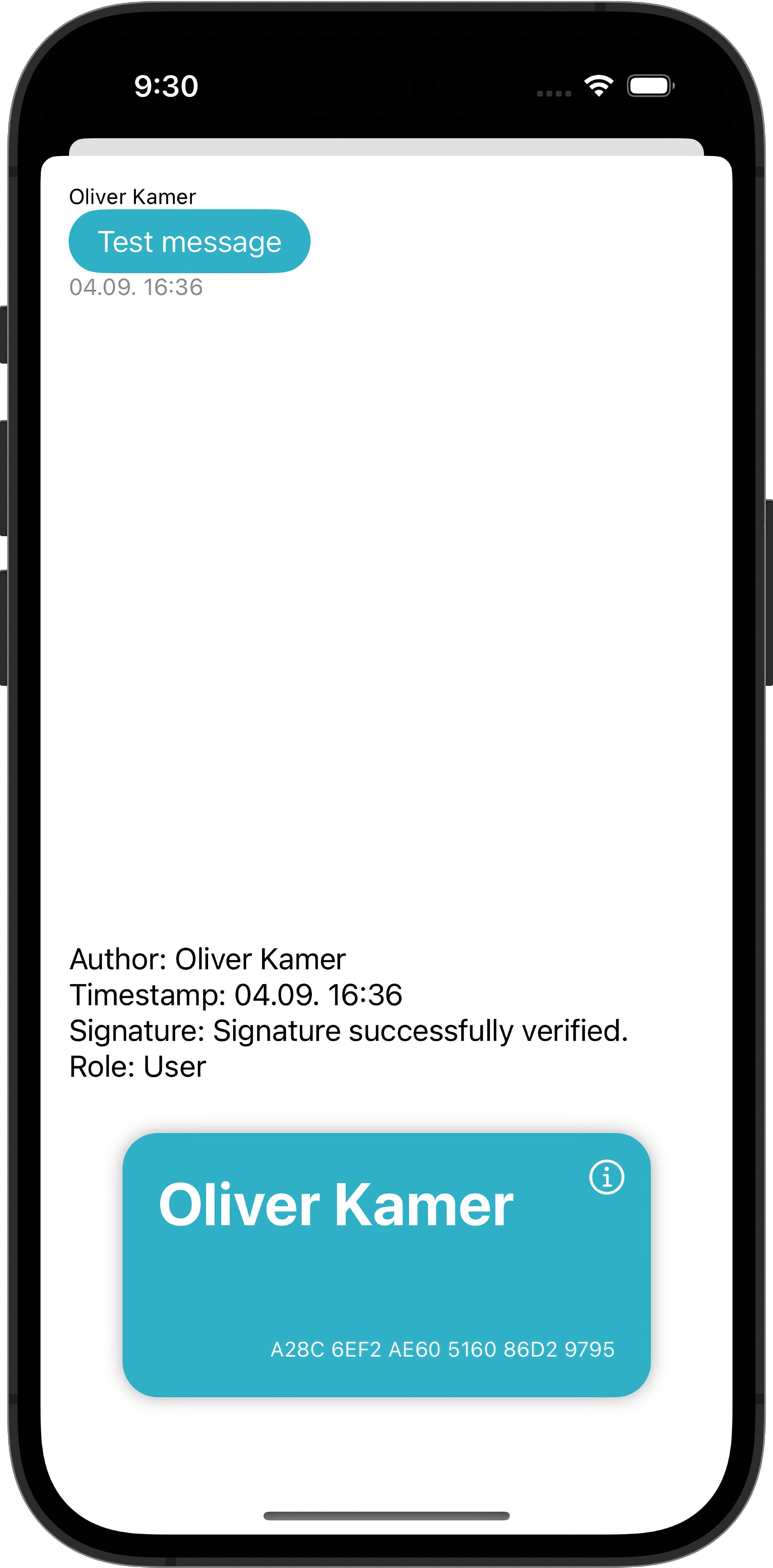}} 
        \subfigure[]{\includegraphics[width=0.185\textwidth]{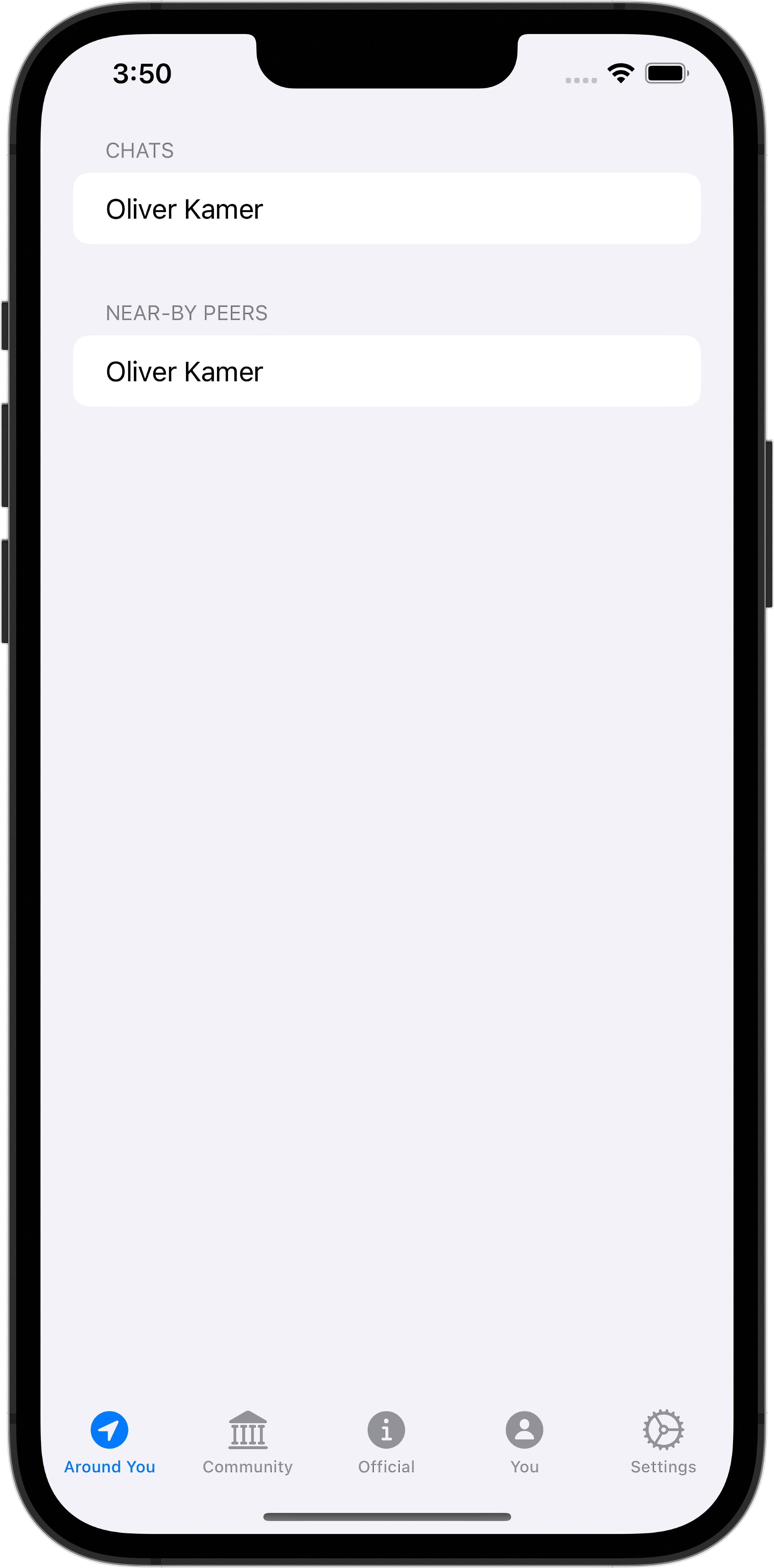}}
        \subfigure[]{\includegraphics[width=0.185\textwidth]{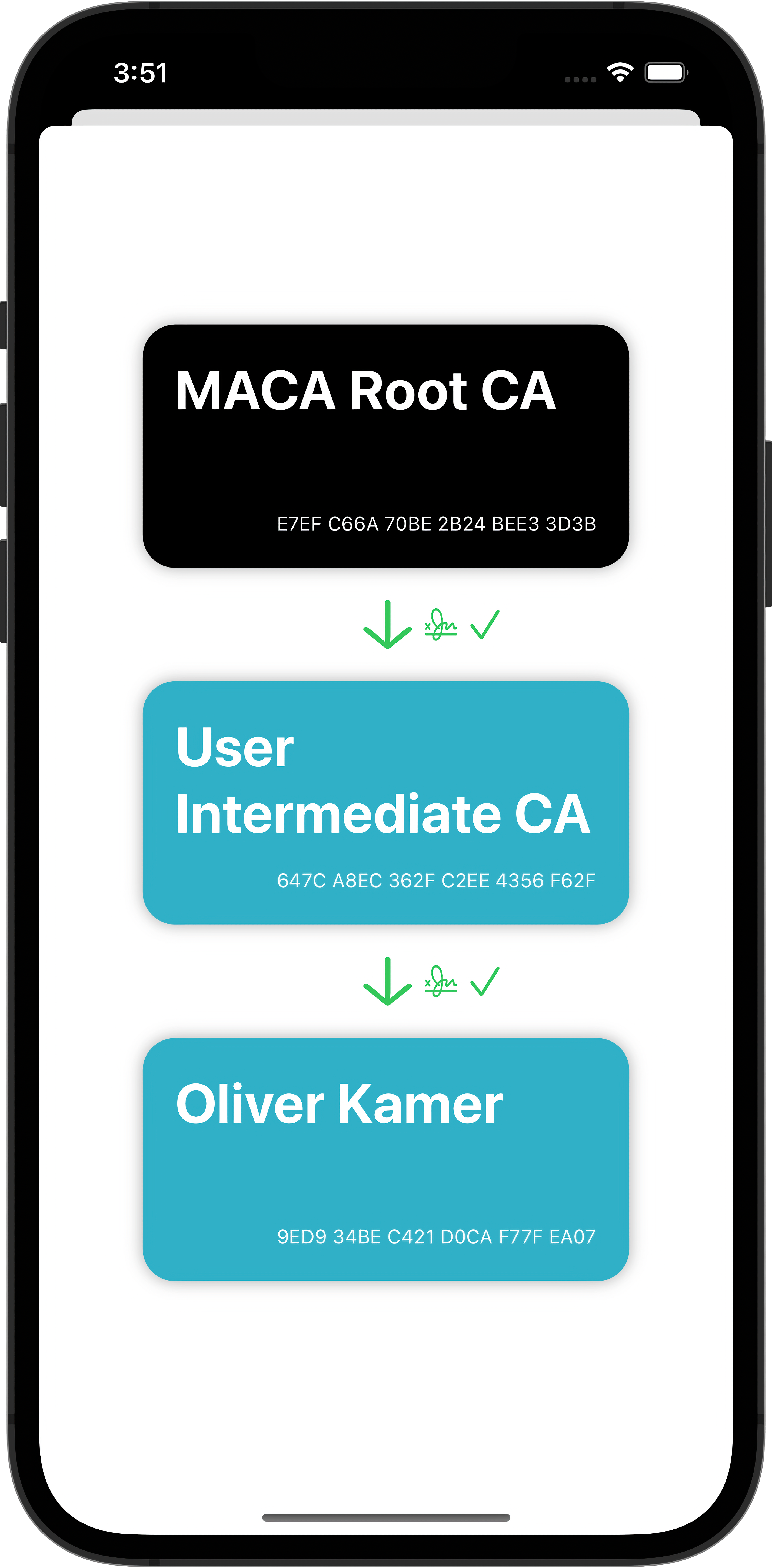}}
        \caption{Messaging service: (a) Community Chat, (b) Message Details, (c) Nearby Users, (d) Certificate Chain}
        \label{fig:community_chat}
\end{figure*}

\subsection{Identity Service}

Anonymity in digital platforms often leads to unwarranted behaviors \cite{zimmerman_online_2016}. Introducing a chat functionality that anchors users to their real-world identities can redefine online communication dynamics. Therefore, we provide an identity service based on a lightweight Public Key Infrastructure (PKI) that encompasses the creation of the digital identity and its application later to sign messages. Once a user submits a government-ID scan, the back-end returns a personal X.509-style certificate. Every message is signed with the corresponding private key, so provenance is verifiable on any device and revocation is immediate. This allows for both reducing misinformation/spamming and enables to create user roles:
\begin{itemize}
    \item Unauthenticated: does not have certificates. May read but not post messages.
    \item Authenticated: holds a personal certificate. Can post signed messages to their location or directly to nearby peers.
    \item Community moderators: authenticated users that carry an additional "moderator" flag. They can flag, hide, or rate-limit content within their zip-code community (\textit{i.e.,} location).
    \item Administrator: can approve or revoke identities, assign moderators, broadcast alerts, and delete messages. 
\end{itemize}

Figure \ref{fig:signup_process} shows the user sign-up process. The user opens the application to a welcome screen, where a private key is securely generated and stored in the keychain. The keychain provides secure storage without requiring custom implementation and enables easy application access. A certificate signing request is created using a third-party package through a simple button press, then transmitted to the backend identity application for system administrator approval.

\subsection{Messaging Service}

After identity creation, the messaging service enables communication using certificates and private/public keys to sign messages. While messaging and identity components are detached (cf. Figure \ref{fig:architecture}), the mobile application requires only messaging functionality once identity setup completes. Users communicate through peer-to-peer messaging (cf. Figure \ref{fig:community_chat} - a) and location-grouped community chat addressed by zip code (cf. Figure \ref{fig:community_chat} - c), supporting both radio and cellular transmission when available. Each message includes date, ID, content, signature, and sender's certificate, as illustrated in Figure \ref{fig:architecture} bottom left.

Message authenticity requires three validations: valid signature, valid certificate with intermediary certificate, and valid certificate resolving through the certificate chain to root certificate. Official messages are protected through two mechanisms: only system administrators can create official message objects, and only messages signed by the official intermediary certificate are valid.

\subsection{Radio Module and Meshtastic}

The physical layer foundation to achieve an off-grid communication. To meet requirements R1 and R2, the system allows nodes to communicate directly without reliance on pre-existing base stations or gateways. In contrast to Narrowband-IoT (NB-IoT) that operates within a licensed cellular spectrum, LoRa operates on ISM frequency bands (\textit{e.g.,} 868 MHz in Europe) being available for spontaneous use \cite{semtechWhatsLoraMain}.

The system leverages the Meshtastic open-source firmware \cite{meshtasticMain}, which provides a complete protocol stack for building a LoRa-based MANET. In addition, it provides  the crucial link between the radio hardware and the mobile application via a protobuf-based API over a BLE link \cite{meshtastic_RadioSettings}, as shown in Figure \ref{fig:architecture}. Meshtastic organizes physical layer settings into "modem presets," which are pre-defined combinations of LoRa parameters that balance the trade-off between data rate and range. Logical communication occurs over "channels," which are encrypted groups for messages. 

\subsection{Field Testing}

\begin{figure*}[ht]
        \centering
        \subfigure[]{\includegraphics[width=0.2\textwidth]{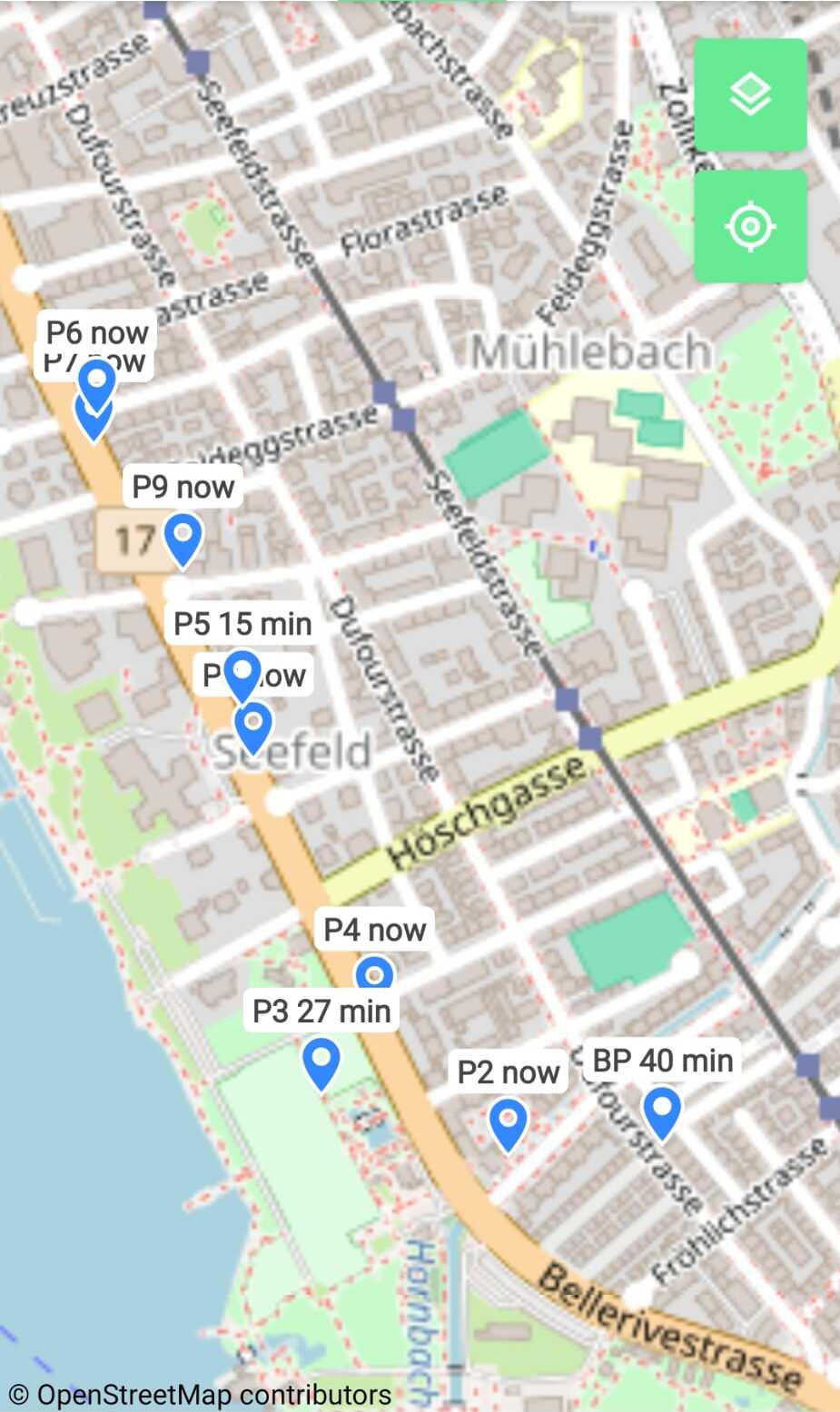}} 
        \subfigure[]{\includegraphics[width=0.2\textwidth]{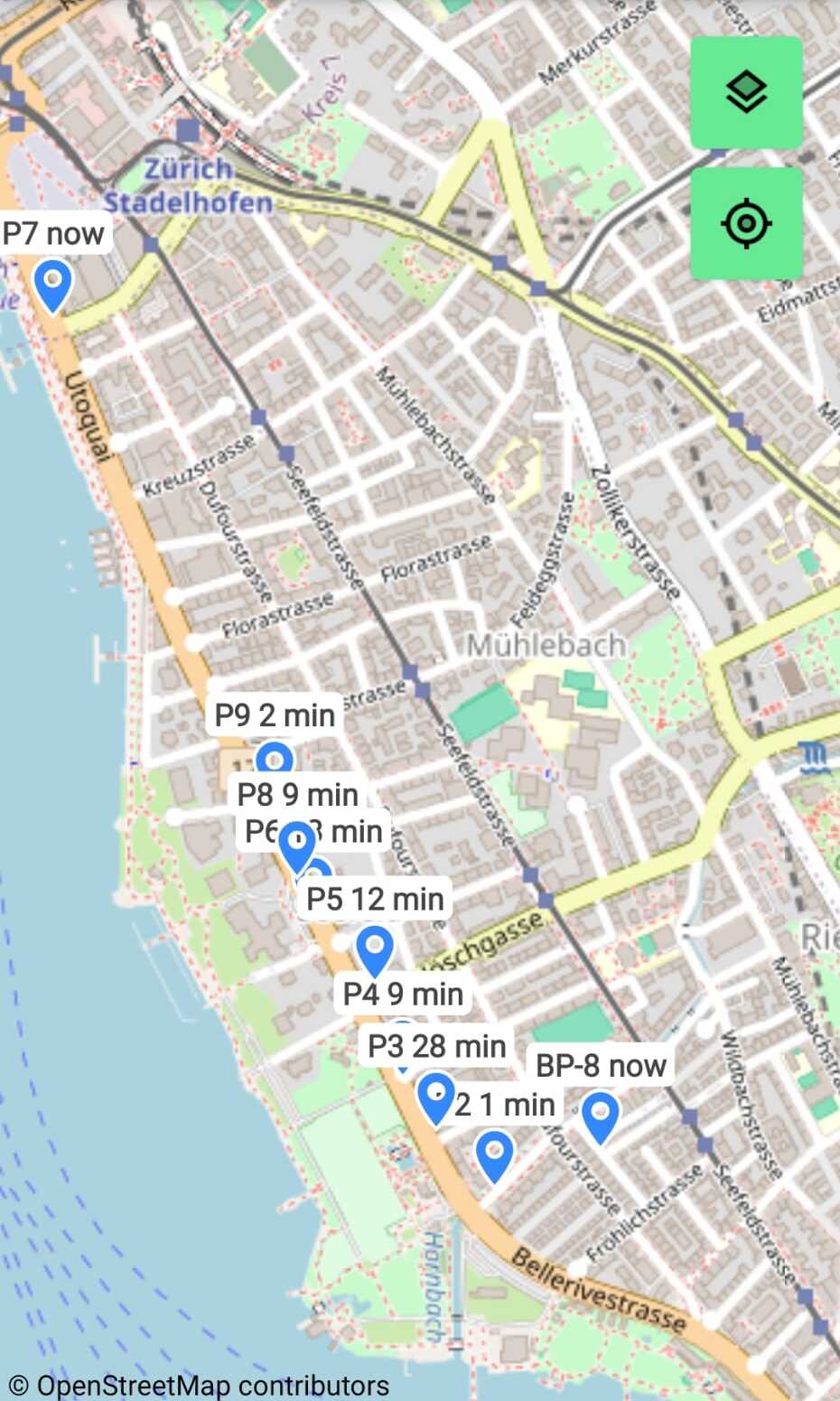}} 
        \subfigure[]{\includegraphics[width=0.2\textwidth]{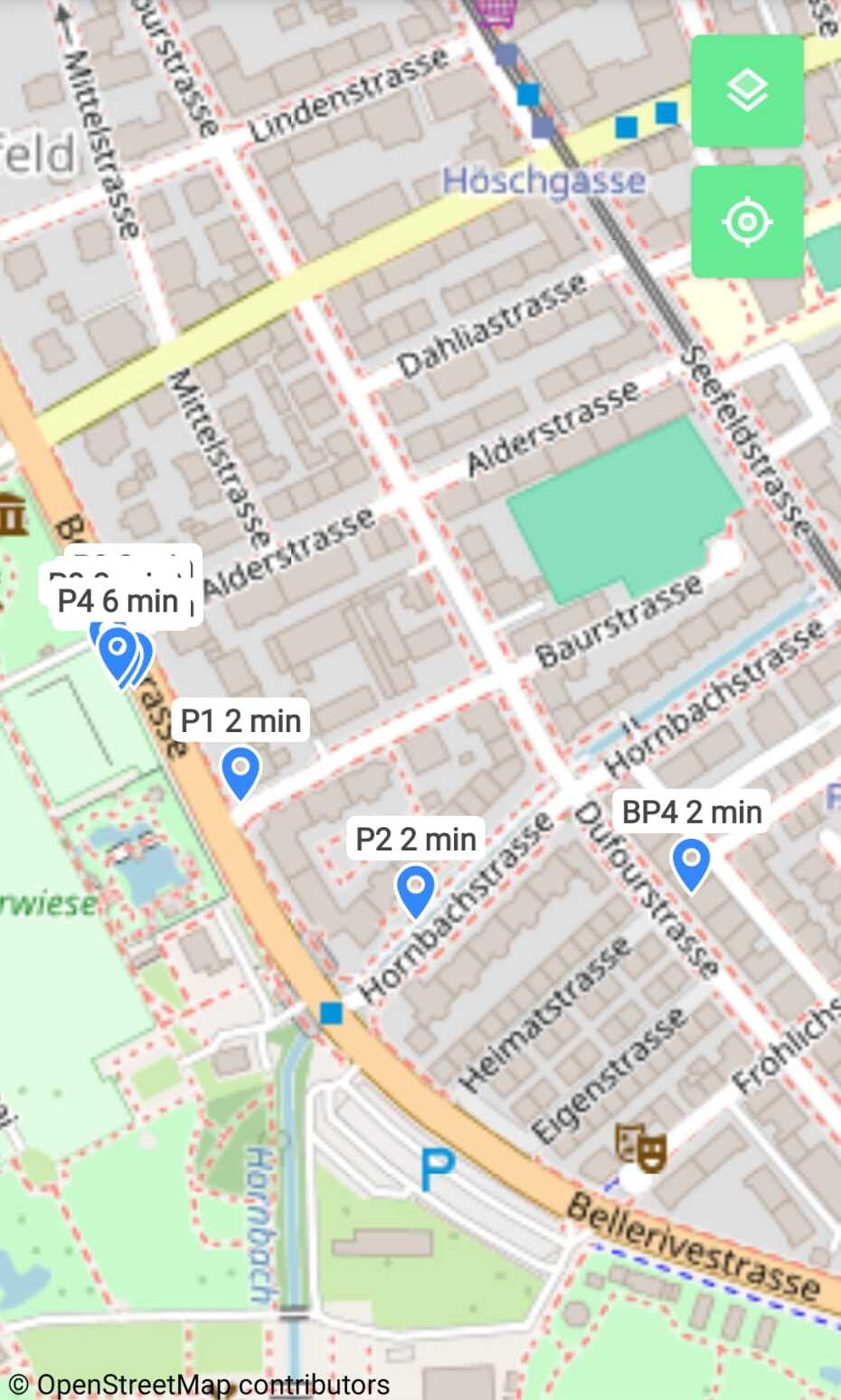}}
        \subfigure[]{\includegraphics[width=0.2\textwidth]{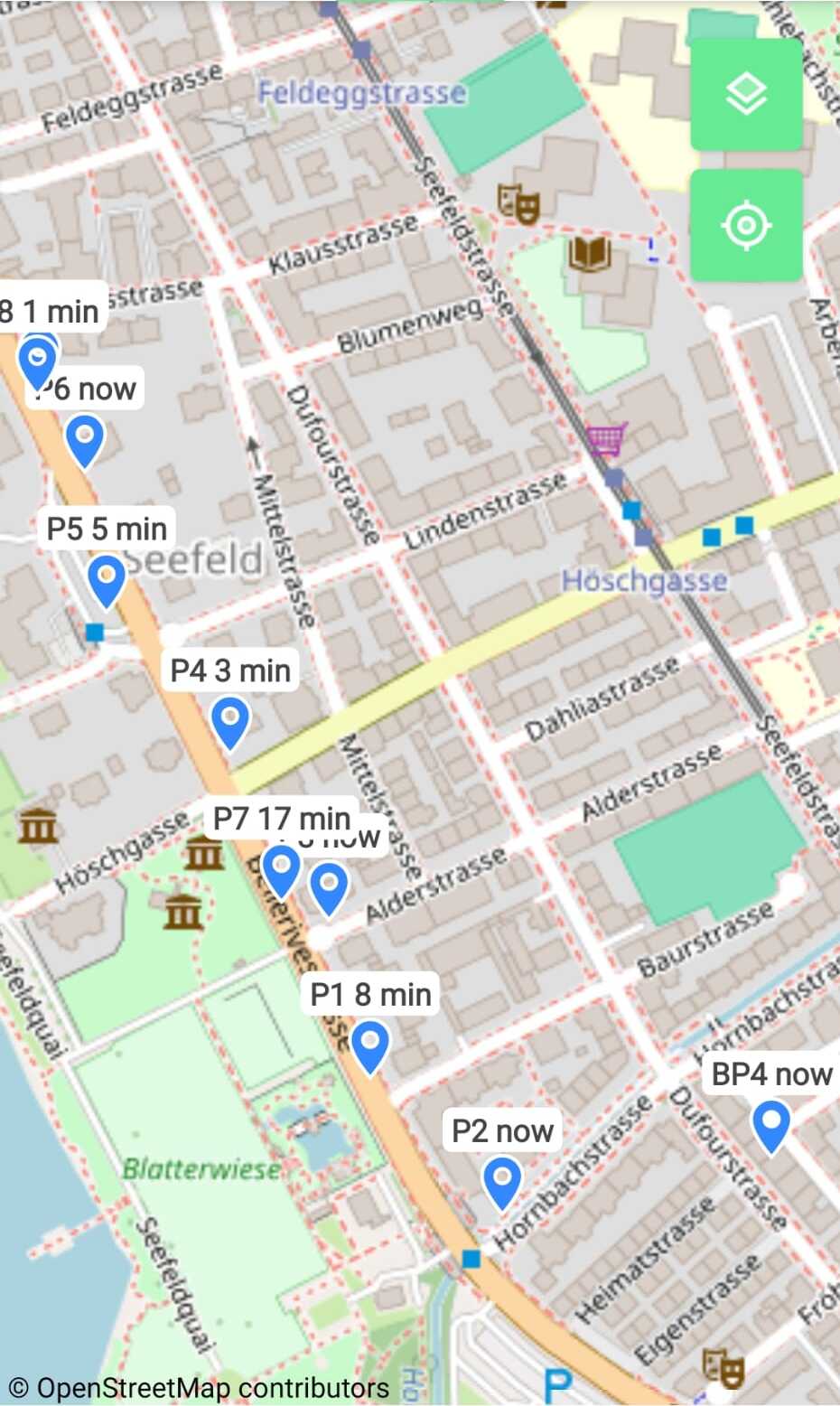}}
        \caption{Nodes on the Map: (a) 868 MHz ShortFast, (b) 868 MHz LongFast, (c) 433 MHz ShortFast, (d) 433 MHz LongFast}
        \label{fig:improvements}
\end{figure*}

The system's performance was evaluated by performing field tests in urband Zürich with nodes deployed at varying distances (100 meters apart, cf. \figurename{} \ref{fig:improvements}), recording PDR, latency, and signal strength metrics. Two test groups were designed and conducted using identical scenarios, locations, and methodology, differing only in utilized radio devices and corresponding frequency range: 868 MHz and 433 MHz.

Each group conducted two field tests using the same frequency. Within each group, two communication channels were tested to determine optimal emergency communication channels. Channel configuration involves different LoRa parameters (Spreading Factor, Coding Rate, Bandwidth), creating trade-offs between data rates, range, and link budget.
\begin{itemize}
    \item The \textbf{Short Range Fast Channel} provides the highest data rate possible. However, it also has the smallest link budget and, thus, a shorter range. 
    \item The \textbf{Long Range Fast Channel} consists of the most balanced LoRa configuration parameters, allowing for a longer range while maintaining a fast data rate.
\end{itemize}

Ten nodes were employed in each field test and split into two groups: sending and receiving nodes. The first node, a central non-moving node that sends packets to other nodes, while the rest are moving nodes that receive messages from the central node. This setup simulates a central contact point for volunteers or community members coordinating rescue operations. 

Field tests utilized the \textit{Range Test} module of Meshtastic, which transmits numbered messages at configurable intervals for performance evaluation. Testing methodology comprised three stages:

\begin{enumerate}
    \item \textbf{Device configuration}: Receiving devices (moving nodes) were configured for continuous monitoring (time interval set to zero). Sending devices (central node) used transmission intervals following Meshtastic recommendations~\cite{MeshtasticRTestConfig}: 30 seconds for LongFast channel, 15 seconds for ShortFast channel.
    \item \textbf{Data collection}: Range tests began after all receiving nodes passed sanity checks and received the first sequence. Distance between sending and receiving nodes increased in 50-meter increments. At 100 meters, one receiving node was fixed to relay messages to further nodes. Testing continued until the farthest receiving node stopped receiving messages. Receiving nodes transmitted only telemetry data to the network.
    \item \textbf{Data extraction and analysis}: Range test results, including message types, telemetry, and location data, were exported from mobile devices in \verb|.csv| format. To avoid memory limitations (50 kB) and ensure data completeness, only the farthest receiving node and sending node were connected to mobile devices for complete data capture.
\end{enumerate}

\section{Results and Evaluation} \label{sec:results}

Two testing group sets were designed, with two subtests in each group. The testing scenario and location are the same for all field tests. The major difference between the two main field tests is the frequency utilized, \textit{i.e}, 868 MHz and 433 MHz frequencies. 

\tablename~\ref{tab:performanceResults} summarizes the results of the acquired metrics from the various field tests, showing the communication channel's mean Received Signal Strength Indication (RSSI) and Signal-to-Noise Ratio (SNR), the achieved maximum distance from the base point, and the overall PDR. 

\begin{table}[H]
\centering
\caption{Performance Metrics by Frequency and Channel}
\resizebox{\columnwidth}{!}{%
\begin{tabular}{llrrrr}
\toprule
Frequency &   Channel &  Mean SNR (dB) &  Mean RSSI (dBm) &  Max Distance (m) &  PDR (\%) \\
\midrule
  868 MHz &  LongFast &           -3.68  &             -113.95 &              1274 &     92.00 \\
  868 MHz & ShortFast &           2.99  &             -121.45  &               786 &     58.57 \\ \hline
  433 MHz &  LongFast &           -1.90  &             -104.42  &               576 &     40.98 \\
  433 MHz & ShortFast &           0.92 &             -84.27 &               281 &     51.06 \\
\bottomrule
\end{tabular}}

\label{tab:performanceResults}
\end{table}

The SNR with the corresponding location data was visualized using the base point data export, while PDR was calculated from the most distant node's data. The distance and payload columns from the data export were used for PDR calculations. The distance column indicated the distance between the sending and receiving nodes, and the payload column contained sequenced messages.

PDR was calculated for distance ranges increasing in 50-meter increments. For example, in the 0–50~m range, all messages within this distance were considered. For each range, PDR was calculated by dividing received messages by total sent messages, where the highest sequence number indicates the total sent. For instance, receiving sequence numbers 1 and 5 from 5 total sent messages (sequences 1-5) yields a PDR of 40\%.

RSSI values were primarily extracted from the base point node, but when packets started hopping (\textit{e.g.} with the 868 MHz LongFast channel at greater distances), RSSI values from other receiving nodes were utilized. 

\begin{figure}[H]
    \centering
    \resizebox{\columnwidth}{!}{\includegraphics{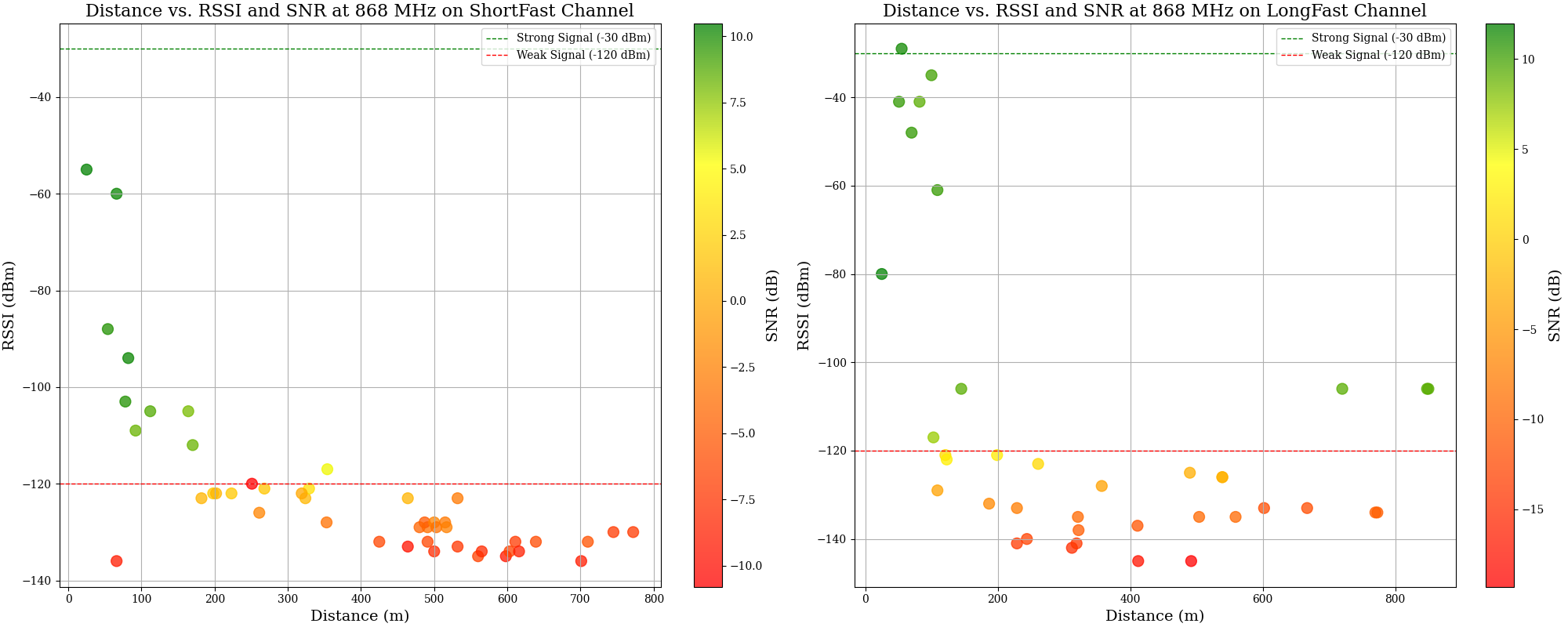}}
    \caption{Distance vs. RSSI and SNR for 868 MHz on ShortFast and LongFast Channels}
    \label{fig:combined_channel_rssi}
\end{figure}

\begin{figure}[H]
    \centering
    \resizebox{\columnwidth}{!}{\includegraphics{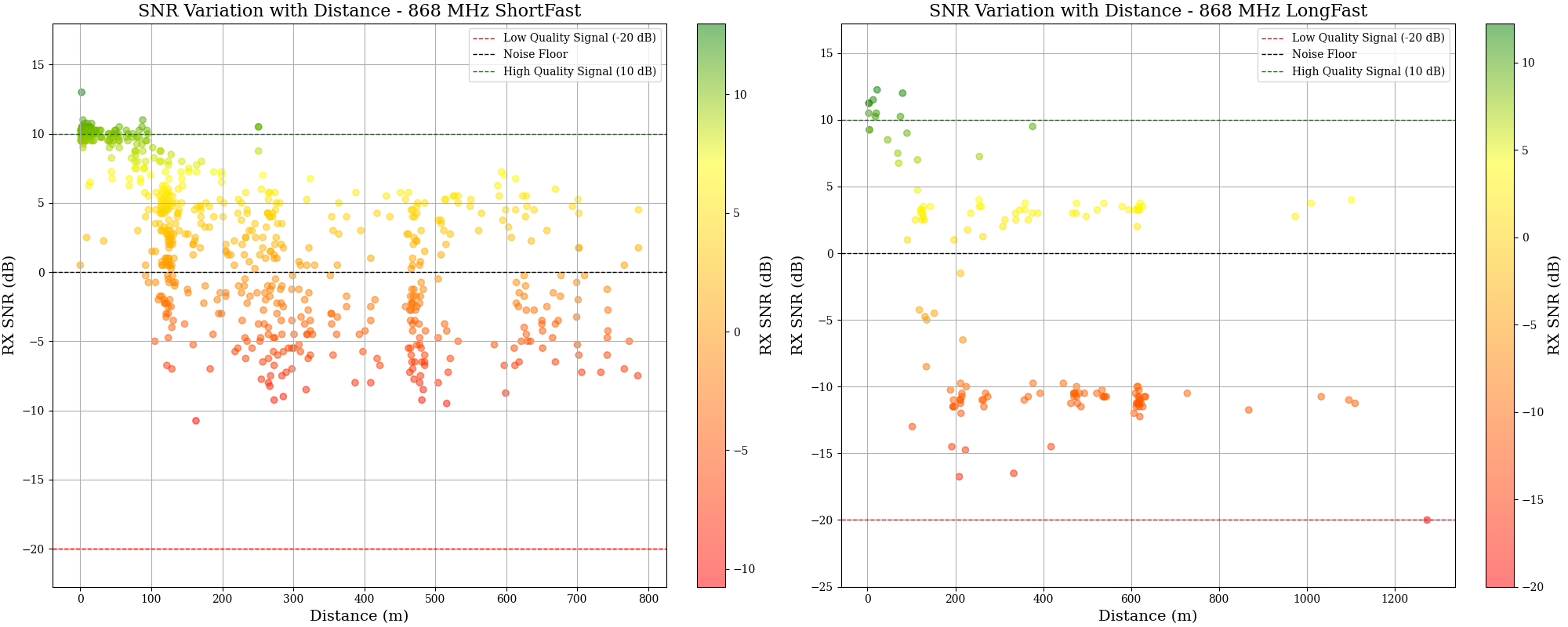}}
    \caption{SNR Variation with Distance for ShortFast and LongFast Channels at 868 MHz}
    \label{fig:snr_distance_868}
\end{figure}

\figurename{} \ref{fig:combined_channel_rssi} and \ref{fig:snr_distance_868} present the RSSI and SNR metrics for field tests on the 868 MHz frequency, comparing the ShortFast (left) and LongFast (right) channels. The first two plots in \figurename{} \ref{fig:combined_channel_rssi} show RSSI and corresponding SNR values over distance. While the LongFast channel achieved greater range, RSSI recordings were limited to approximately 850 m due to the multi-hopping mechanism of the routing protocol, which drops certain information when no direct link exists. Although RSSI is captured within SNR calculations, the two metrics provide distinct insights. Signal strength diminishes beyond 200 m on both channels, with RSSI values generally ranging between -120 dBm and -140 dBm, indicative of weak signals. Up to 800 m, both channels exhibit similar RSSI values.

\figurename{} \ref{fig:snr_distance_868} compares SNR values for both channels. The ShortFast channel shows better signal quality with most values above -10 dB, while the LongFast channel records several values near -20 dB, even at 200 m. This difference results from higher external node density on the LongFast channel (30+ devices) versus ShortFast (10 devices).

The ShortFast channel's 15-s transmission interval produced twice as many messages as LongFast, explaining its higher data record count. Lower ShortFast PDR delayed distance increments until successful message reception, enabling telemetry logging to the base point

\subsection{Radio Frequency: 868 vs. 433 MHz} 

\begin{figure}[h]
    \centering
    \resizebox{\columnwidth}{!}{\includegraphics{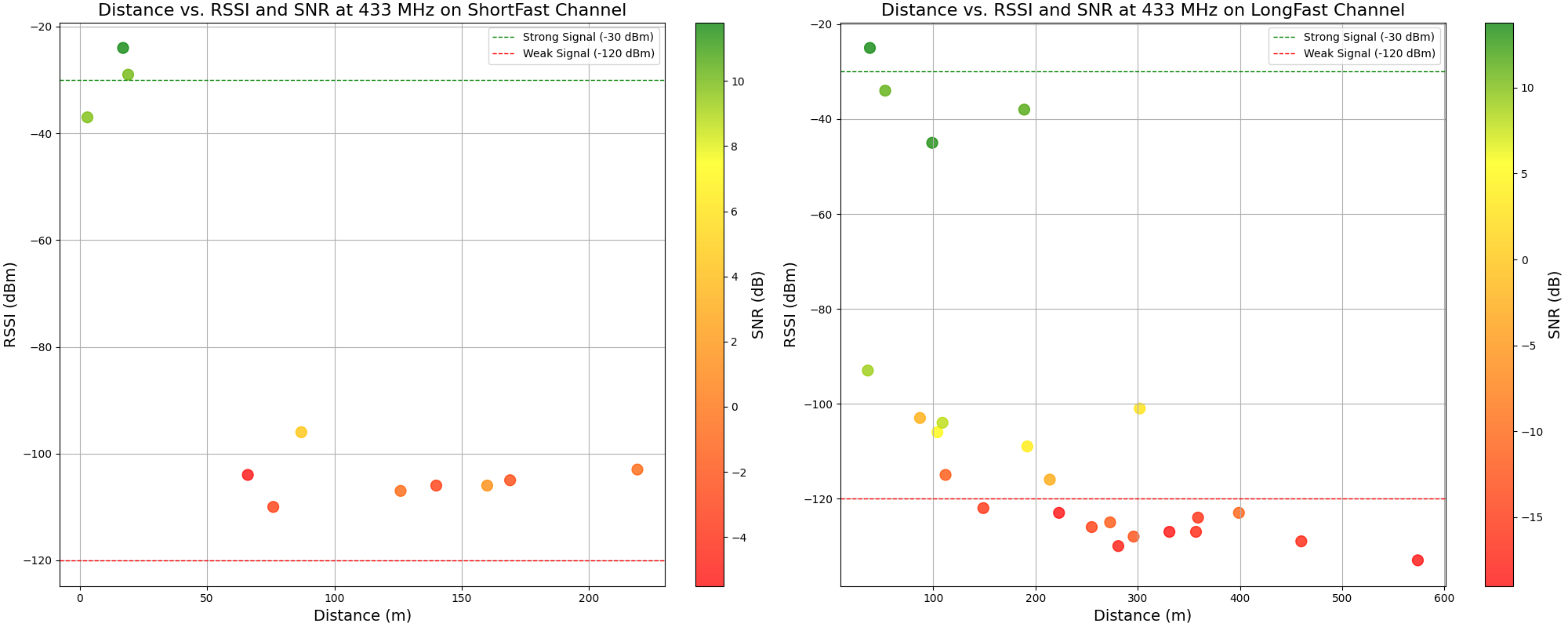}}
    \caption{Distance vs. RSSI and SNR for 433 MHz on ShortFast and LongFast Channels}
    \label{fig:combined_channel_rssi2}
\end{figure}

\begin{figure}[h]
    \centering
    \resizebox{\columnwidth}{!}{\includegraphics{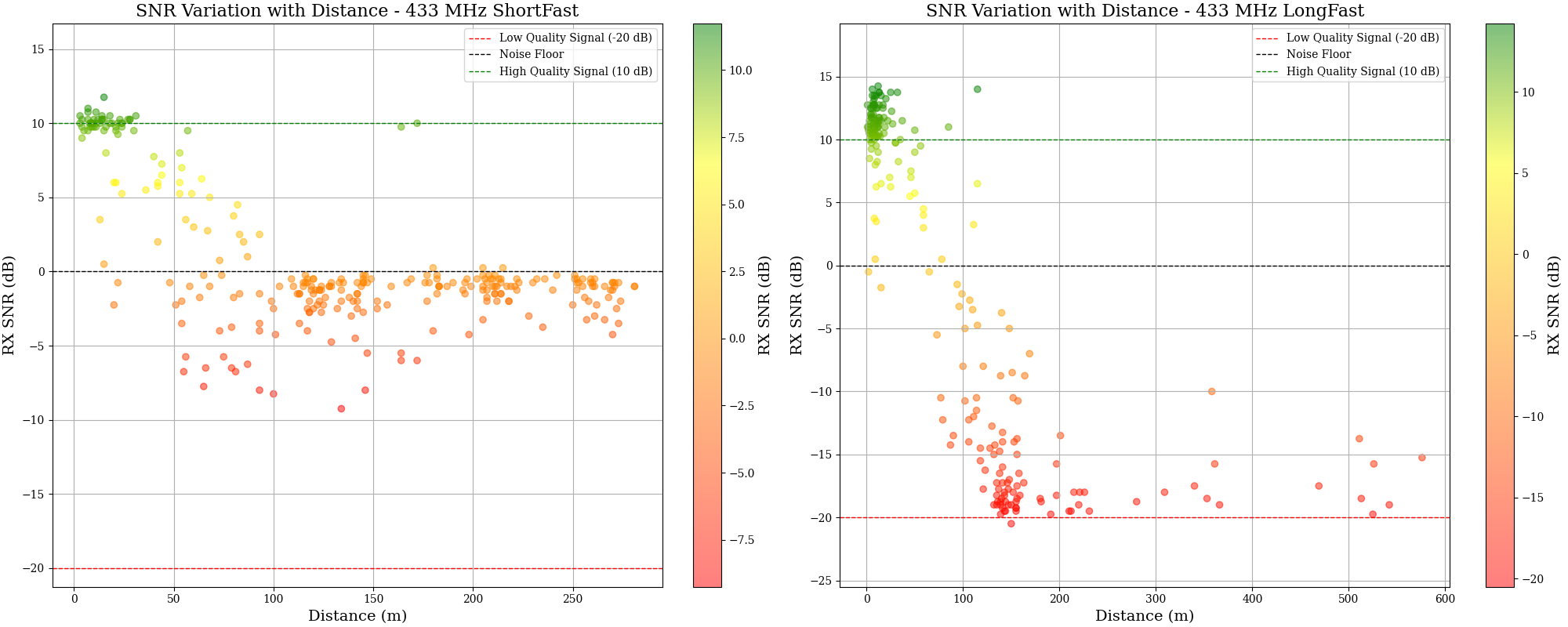}}
    \caption{SNR Variation with Distance for ShortFast and LongFast Channels at 433 MHz}
    \label{fig:snr_distance}
\end{figure}

\figurename~\ref{fig:combined_channel_rssi2} and \ref{fig:snr_distance} present the RSSI and SNR values over distance for the 433 MHz frequency on both ShortFast channel (left) and LongFast channel (right). \figurename~ \ref{fig:combined_channel_rssi2} compares RSSI values for the ShortFast and LongFast channels. All recorded signals on the ShortFast channel remained above the weak signal threshold, exhibiting stronger signals than those on the LongFast channel. However, the ShortFast channel achieved a shorter maximum distance.

\figurename~\ref{fig:snr_distance} highlights SNR variations over distance for both channels. Similar to RSSI, SNR values on the ShortFast channel were generally far from the low-quality threshold, while the LongFast channel recorded several signals exceeding the -20 dB level. As expected, the LongFast channel achieved a greater range but at the cost of weaker signal quality. The ShortFast channel provided better signal strength and quality, while the LongFast channel excelled in distance, achieving nearly double the range of the ShortFast channel.

\begin{figure}[h]
    \centering
    \resizebox{\columnwidth}{!}{\includegraphics{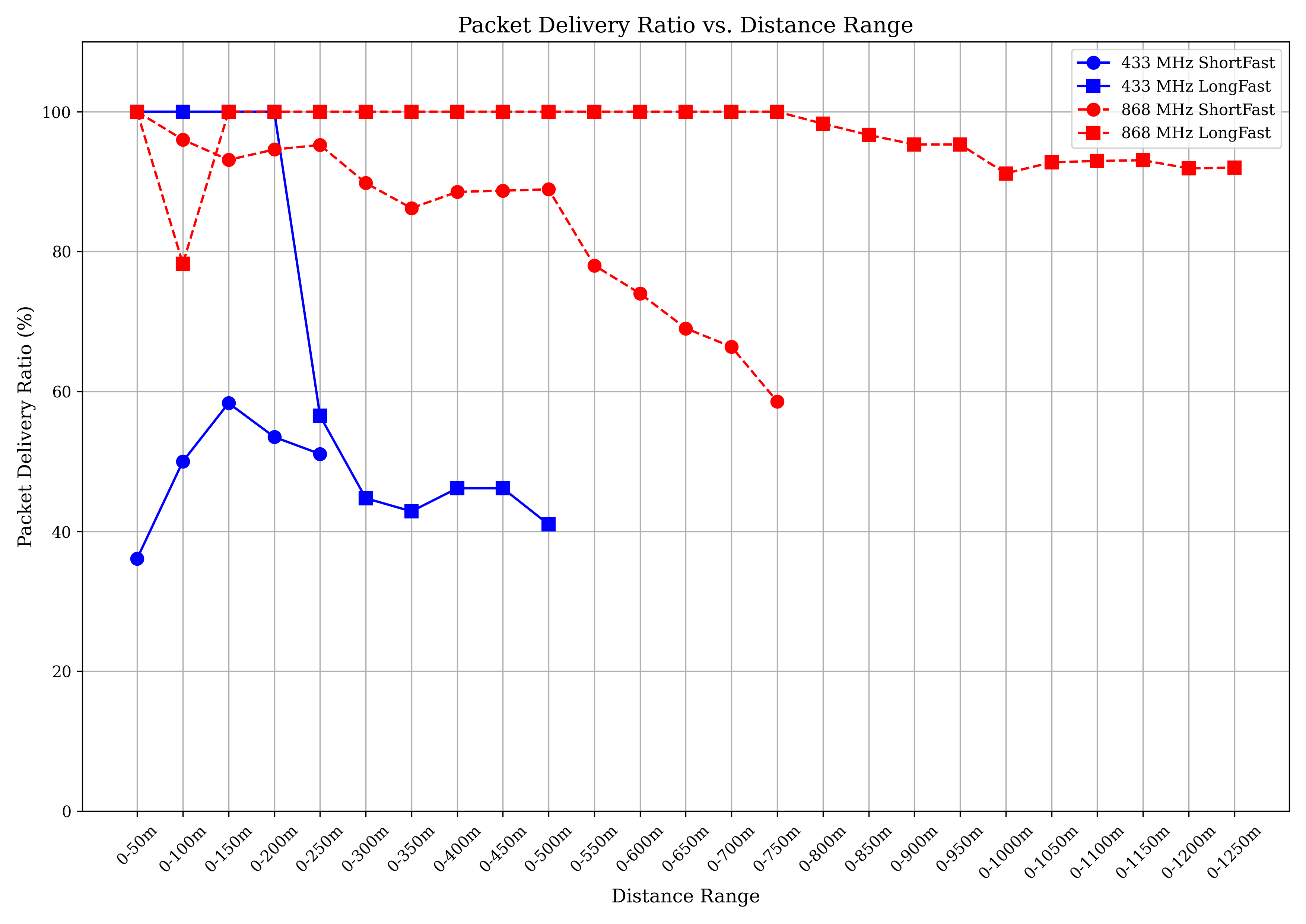}}
    \caption{Packet Delivery Ratio vs. Distance Range}
    \label{fig:pdr_vs_distance}
\end{figure}

\figurename~\ref{fig:pdr_vs_distance} compares PDR values and achieved distances for both frequency ranges (868 MHz in red and 433 MHz in blue). The results highlight the superior performance of the 868 MHz frequency, which achieves both higher PDR values and greater range. Notably, the 868 MHz ShortFast channel outperforms the 433 MHz LongFast channel, even though the latter is designed for extended range.

All channels exhibit an abrupt drop in PDR to zero at their maximum range rather than a gradual decline. For example, the 868 MHz LongFast channel maintains a PDR of approximately 85 up to 1,250 m before dropping to zero. To ensure accuracy, each test allowed sufficient time for 10 messages to be sent before concluding. For example, on the LongFast channel with a 30-second message interval, tests were terminated after five minutes of no message reception, and the maximum attainable distance was recorded.

\subsection{Mobile Application}
To evaluate the effectiveness of the mobile application component for civilian emergency communication, a requirement-based analysis was conducted. The evaluation seen in Table \ref{tab:req_evaluation_mobile} focused on requirements R2, R4, R5, R6, and R7, which target user accessibility and interface design challenges.

\begin{table}[h]
\centering
\caption{Mobile Application of Requirements}
\label{tab:req_evaluation_mobile}
\resizebox{\columnwidth}{!}{%
    \begin{tabular}{c c p{6cm}}
        \toprule
        \textbf{ID} & \textbf{Eval.} & \textbf{Description} \\
        \midrule
        R1 & \cmark    & Achieved 1.2 km range and 92\% PDR in urban field tests. \\
        R2 & \cmark    & BLE bridge abstracts radio complexity for phone use. \\
        \addlinespace
        R3 & \feature1 & Low-power hardware used, but 24h endurance was not tested. \\
        \addlinespace
        R4 & \feature1 & High usability validated, but prototype was iOS-only. \\
        \addlinespace
        R5 & \cmark    & Implemented a full PKI system with verified identity and roles. \\
        \addlinespace
        R6 & \cmark    & System design supports automatic switching between local P2P and cellular. \\
        \addlinespace
        R7 & \feature1 & Streamlined setup, but requires manual admin approval for identity. \\
        \bottomrule
    \end{tabular}%
}
    \begin{tablenotes}
    \item $\CIRCLE=\text{property }$; $\LEFTcircle=\text{property partially provided}$;
    $\text{\xmark}=\text{property not provided}$
    \end{tablenotes} 
\end{table}

The 868~MHz LongFast channel, achieving a 1.2~km range with 92\% PDR, provides a reliable physical layer (R1). This contrasts with other off-grid solutions \cite{lowCostMessages, disasteradioMain, lochaMesh, LOCATE_Main}, which are often limited by a shorter range and higher power consumption. Also, most related work lack an authentication and role system and, as such, are open systems for anyone to join and, therefore, not suitable for a volunteer chat application, as this means that anyone can participate and bad actors have no responsibility through their identity that is linked with them \cite{sharevski_beyond_2020}.

The mobile application bridges R2 by enabling users to access LoRa mesh capabilities without requiring an understanding of the underlying radio protocols. The BLE bridge implementation allows civilians to connect mobile phones to radio boards using familiar smartphone interactions for mesh network communication. Cross-platform usability (R4) was validated through system-wide communication across various mobile platforms utilizing a shared LoRa infrastructure. Messages transmitted through the LoRa mesh network are received and processed regardless of mobile platform. The application interface utilizes established messaging conventions to ensure intuitive operation during emergency scenarios.

For verified identity and moderation requirement (R5), the application's integrated PKI system enables cryptographic verification in a user-friendly format, allowing users to verify message authenticity through signature validation. Furthermore, the role-based moderation system enables community moderators to maintain communication quality. The cellular fallback capability (R6) serves as an intelligent switching mechanism within the application. The application automatically selects the most suitable communication path based on availability, eliminating the need for user intervention.

Finally, the rapid bootstrap requirement (R7) was validated through simplified emergency deployment. Once users complete the initial identity verification, the system becomes immediately operable after establishing Bluetooth connectivity with the LoRa device. Unauthenticated users access the system in read-only mode, receiving critical emergency information without setup requirements.

\vspace{-0.5em}
\subsection{Discussion}

The primary goal of the field experiments was to evaluate whether a \textbf{LoRa-based MANET} can effectively function in emergencies and determine the most suitable frequency (868 MHz vs. 433 MHz) and channel configuration parameters in Zürich. 

LoRa configuration parameters were grouped by channels, balancing range, data rate, and link budget. While LoRa excels in long-range communication, achieving high values for all three metrics is challenging, particularly in urban environments with significant signal-blocking objects. Nonetheless, LoRa's affordability, high link budget, and ability to handle low-quality signals make it a viable alternative in emergencies.

The 868 MHz frequency was superior in the Zürich urban setting, outperforming 433 MHz in range, PDR, and signal quality. For instance, the 868 MHz LongFast channel maintained nearly 10 dB better signal quality at twice the distance (600 m vs. 300 m) compared to the 433 MHz LongFast channel. This superiority is partly due to higher transmission power (25 mW vs. 10 mW) and reduced interference from other devices, which are more common on the 433 MHz frequency. While both frequencies adhered to duty cycle limits, no significant issues arose during the tests. Messages' size was small (\textit{e.g.,} \verb|"seq 5"|), and the ShortFast channel maintained 15-second intervals without exceeding utilization limits. Power consumption was moderate, with nodes consuming 2–10\% battery during one-hour tests, depending on Bluetooth connectivity. Devices on 868 MHz, despite higher power output, showed no notable difference in power usage compared to 433 MHz.


The \textbf{mobile application} evaluation shows that cryptographic operations can be integrated into civilian-accessible interfaces, according to best practices listed in \cite{preneel_cryptography_2011}. The requirement-based analysis demonstrates that technical resilience and user accessibility can coexist within emergency communication systems. The PKI implementation enables message authenticity verification while maintaining intuitive operation, addressing the critical challenge of preventing misinformation during crisis scenarios. Additionally, cross-platform compatibility through shared LoRa infrastructure validates the system's deployment flexibility across diverse device ecosystems. This approach eliminates platform-specific constraints that have limited previous emergency communication solutions. The cellular fallback mechanism enables intelligent adaptation to varying infrastructure conditions, ensuring continuous communication as network availability fluctuates during emergencies.

The unified approach of combining validated LoRa mesh performance with accessible smartphone interfaces addresses the gap in emergency communications between technical capability and civilian adoption. Unlike previous work that treats these dimensions separately, our system demonstrates that both requirements can be satisfied simultaneously without compromising either technical performance or user accessibility.

\vspace{-0.3em}

\subsection{Limitations}





First, real-world conditions and logistical challenges constrained the field tests. Weather dependency limited testing to favorable conditions, potentially overlooking performance degradation during adverse weather scenarios typical in emergencies. Coordinating participants within short timeframes was difficult, resulting in device placement at fixed locations rather than the intended mobile mesh network simulation, where users carry the devices.

Second, the mobile application evaluation was conducted as a separate component study, rather than as part of integrated system testing. While requirement-based analysis validates component compatibility, actual integration performance remains to be validated through comprehensive end-to-end testing. Furthermore, the identity verification process assumes the availability of government ID validation infrastructure, which may not be present in all deployment contexts. Alternative verification processes are therefore necessary to maintain security while ensuring broader accessibility.

Third, the network scalability assessment was limited to 10 nodes, which may not accurately reflect the performance characteristics of larger emergency networks communicating over the same LoRa channel. Additionally, while the field tests were conducted in urban Zürich, other urban testing environments with different geographic or demographic characteristics may show different results in terms of performance. Building densities, terrain topography, and electromagnetic interference patterns can all impact the signal and network reliability. 

Finally, the mobile application dependency on pre-emergency identity setup creates barriers for spontaneous network participation. While unauthenticated users can receive messages, the inability to send messages without prior verification may limit the effectiveness of community-driven emergency responses.

\section{Summary and Future Work} \label{sec:conclusion}

This paper presented a unified emergency communication system integrating LoRa-based MANET infrastructure with a civilian accessible mobile application. Field experiments in urban Zürich validated the performance of the LoRa mesh network. The 868 MHz frequency, using the LongFast configuration, achieved optimal performance (1.2 km range and 92\% PDR), significantly outperforming the 433 MHz frequency. These results establish the viability of LoRa for urban emergency communications as a potential network infrastructure. The mobile application component demonstrated that emergency communication features can be made accessible to civilian users. Requirement-based evaluation confirmed the mobile application achieves cross-platform usability, PKI-based identity verification, and rapid deployment without compromising technical capabilities.

Future work will explore additional LoRa configuration parameters (like power consumption and channel utilization) through expanded field tests with larger node counts (50+ nodes) and diverse geographic environments (rural and urban areas). Comparisons with alternative communication systems and comprehensive end-to-end validation remain crucial for practical deployment. Lastly, exploring integration possibilities with existing emergency response infrastructures is important. This, for example, will explore how to make the system more interoperable and directly coordinate and sync the civilian network with professional emergency services.

\begin{acks}
This work builds upon Karim's Master Thesis submitted at UZH, contributing to the CYREN Project, while extending gratitude to Melanie Knieps for her invaluable support during the field tests and usability questionnaires.
\end{acks}

\bibliographystyle{ACM-Reference-Format}
\bibliography{bib/sample-base}


\begin{thebibliography}{30}


\ifx \showCODEN    \undefined \def \showCODEN     #1{\unskip}     \fi
\ifx \showISBNx    \undefined \def \showISBNx     #1{\unskip}     \fi
\ifx \showISBNxiii \undefined \def \showISBNxiii  #1{\unskip}     \fi
\ifx \showISSN     \undefined \def \showISSN      #1{\unskip}     \fi
\ifx \showLCCN     \undefined \def \showLCCN      #1{\unskip}     \fi
\ifx \shownote     \undefined \def \shownote      #1{#1}          \fi
\ifx \showarticletitle \undefined \def \showarticletitle #1{#1}   \fi
\ifx \showURL      \undefined \def \showURL       {\relax}        \fi
\providecommand\bibfield[2]{#2}
\providecommand\bibinfo[2]{#2}
\providecommand\natexlab[1]{#1}
\providecommand\showeprint[2][]{arXiv:#2}

\bibitem[Albrecht et~al\mbox{.}(2021)]%
        {paterson_mesh_2021}
\bibfield{author}{\bibinfo{person}{Martin~R. Albrecht}, \bibinfo{person}{Jorge Blasco}, \bibinfo{person}{Rikke~Bjerg Jensen}, {and} \bibinfo{person}{Lenka Mareková}.} \bibinfo{year}{2021}\natexlab{}.
\newblock \showarticletitle{Mesh {Messaging} in {Large}-{Scale} {Protests}: {Breaking} {Bridgefy}}.
\newblock In \bibinfo{booktitle}{\emph{Topics in {Cryptology} – {CT}-{RSA} 2021}}, \bibfield{editor}{\bibinfo{person}{Kenneth~G. Paterson}} (Ed.). Vol.~\bibinfo{volume}{12704}. \bibinfo{publisher}{Springer International Publishing}, \bibinfo{address}{Cham}, \bibinfo{pages}{375--398}.
\newblock
\showISBNx{978-3-030-75538-6 978-3-030-75539-3}
\href{https://doi.org/10.1007/978-3-030-75539-3_16}{doi:\nolinkurl{10.1007/978-3-030-75539-3_16}}
\newblock
\shownote{Series Title: Lecture Notes in Computer Science}.


\bibitem[{Bridgefy, Inc.}(2023)]%
        {bridgefy_inc_bridgefy_2023}
\bibfield{author}{\bibinfo{person}{{Bridgefy, Inc.}}} \bibinfo{year}{2023}\natexlab{}.
\newblock \bibinfo{title}{Bridgefy – {Offline} {Messages} {App} \& {SDK}}.
\newblock
\urldef\tempurl%
\url{https://bridgefy.me/}
\showURL{%
\tempurl}


\bibitem[Cardenas et~al\mbox{.}(2020)]%
        {lowCostMessages}
\bibfield{author}{\bibinfo{person}{Angelica~Moreno Cardenas}, \bibinfo{person}{Miguel~Kiyoshy Nakamura~Pinto}, \bibinfo{person}{Ermanno Pietrosemoli}, \bibinfo{person}{Marco Zennaro}, \bibinfo{person}{Marco Rainone}, {and} \bibinfo{person}{Pietro Manzoni}.} \bibinfo{year}{2020}\natexlab{}.
\newblock \showarticletitle{A low-cost and low-power messaging system based on the LoRa wireless technology}.
\newblock \bibinfo{journal}{\emph{Mobile networks and applications}}  \bibinfo{volume}{25} (\bibinfo{year}{2020}), \bibinfo{pages}{961--968}.
\newblock


\bibitem[Deshmukh and Kakarwal(2019)]%
        {dey_1_2019}
\bibfield{author}{\bibinfo{person}{Manjusha Deshmukh} {and} \bibinfo{person}{Sangeeta Kakarwal}.} \bibinfo{year}{2019}\natexlab{}.
\newblock \showarticletitle{1. {Adaptive} routing for emergency communication via {MANET}}.
\newblock In \bibinfo{booktitle}{\emph{The {Internet} of {Everything}}}, \bibfield{editor}{\bibinfo{person}{Nilanjan Dey}, \bibinfo{person}{Gitanjali Shinde}, \bibinfo{person}{Parikshit Mahalle}, {and} \bibinfo{person}{Henning Olesen}} (Eds.). \bibinfo{publisher}{De Gruyter}, \bibinfo{pages}{5--28}.
\newblock
\showISBNx{978-3-11-062851-7}
\href{https://doi.org/10.1515/9783110628517-002}{doi:\nolinkurl{10.1515/9783110628517-002}}


\bibitem[DiFranco(2023)]%
        {difranco_airchat_2023}
\bibfield{author}{\bibinfo{person}{Mark DiFranco}.} \bibinfo{year}{2023}\natexlab{}.
\newblock \bibinfo{title}{{AirChat}: {Peer}-to-{Peer} {Chat}}.
\newblock
\urldef\tempurl%
\url{https://apps.apple.com/us/app/airchat-peer-to-peer-chat/id1606916296}
\showURL{%
\tempurl}


\bibitem[Disasteradio(2024)]%
        {disasteradioMain}
\bibfield{author}{\bibinfo{person}{Disasteradio}.} \bibinfo{year}{2024}\natexlab{}.
\newblock \bibinfo{title}{A disaster-resilient communications network powered by the sun}.
\newblock
\urldef\tempurl%
\url{https://disaster.radio/}
\showURL{%
\tempurl}


\bibitem[for Civil Protection~FOCP(2024)]%
        {alertSwissApp}
\bibfield{author}{\bibinfo{person}{Federal~Office for Civil Protection~FOCP}.} \bibinfo{year}{2024}\natexlab{}.
\newblock \bibinfo{title}{AlertSwiss}.
\newblock
\urldef\tempurl%
\url{https://www.alert.swiss/}
\showURL{%
\tempurl}


\bibitem[Hern(2014)]%
        {hern_firechat_2014}
\bibfield{author}{\bibinfo{person}{Alex Hern}.} \bibinfo{year}{2014}\natexlab{}.
\newblock \showarticletitle{Firechat updates as 40,000 {Iraqis} download 'mesh' chat app in censored {Baghdad}}.
\newblock \bibinfo{journal}{\emph{The Guardian}} (\bibinfo{date}{June} \bibinfo{year}{2014}).
\newblock
\showISSN{0261-3077}
\urldef\tempurl%
\url{https://www.theguardian.com/technology/2014/jun/24/firechat-updates-as-40000-iraqis-download-mesh-chat-app-to-get-online-in-censored-baghdad}
\showURL{%
\tempurl}


\bibitem[Hoebeke et~al\mbox{.}(2004)]%
        {manetChallenges2004}
\bibfield{author}{\bibinfo{person}{Jeroen Hoebeke}, \bibinfo{person}{Ingrid Moerman}, \bibinfo{person}{Bart Dhoedt}, {and} \bibinfo{person}{Piet Demeester}.} \bibinfo{year}{2004}\natexlab{}.
\newblock \showarticletitle{An overview of mobile ad hoc networks: applications and challenges}.
\newblock \bibinfo{journal}{\emph{Journal-Communications Network}} \bibinfo{volume}{3}, \bibinfo{number}{3} (\bibinfo{year}{2004}), \bibinfo{pages}{60--66}.
\newblock


\bibitem[{HypeLabs Inc.}(2020)]%
        {hypelabs_inc_hypelabs_2020}
\bibfield{author}{\bibinfo{person}{{HypeLabs Inc.}}} \bibinfo{year}{2020}\natexlab{}.
\newblock \bibinfo{title}{{HypeLabs}}.
\newblock
\urldef\tempurl%
\url{https://hypelabs.io/}
\showURL{%
\tempurl}


\bibitem[Jang et~al\mbox{.}(2009)]%
        {jang_rescue_2009}
\bibfield{author}{\bibinfo{person}{Hung-Chin Jang}, \bibinfo{person}{Yao-Nan Lien}, {and} \bibinfo{person}{Tzu-Chieh Tsai}.} \bibinfo{year}{2009}\natexlab{}.
\newblock \showarticletitle{Rescue information system for earthquake disasters based on {MANET} emergency communication platform}. In \bibinfo{booktitle}{\emph{Proceedings of the 2009 {International} {Conference} on {Wireless} {Communications} and {Mobile} {Computing}: {Connecting} the {World} {Wirelessly}}}. \bibinfo{publisher}{ACM}, \bibinfo{address}{Leipzig Germany}, \bibinfo{pages}{623--627}.
\newblock
\showISBNx{978-1-60558-569-7}
\href{https://doi.org/10.1145/1582379.1582514}{doi:\nolinkurl{10.1145/1582379.1582514}}


\bibitem[Lien et~al\mbox{.}(2009)]%
        {2009manetHelpEmergency}
\bibfield{author}{\bibinfo{person}{Yao-Nan Lien}, \bibinfo{person}{Hung-Chin Jang}, {and} \bibinfo{person}{Tzu-Chieh Tsai}.} \bibinfo{year}{2009}\natexlab{}.
\newblock \showarticletitle{A MANET based emergency communication and information system for catastrophic natural disasters}. In \bibinfo{booktitle}{\emph{2009 29th IEEE international conference on distributed computing systems workshops}}. IEEE, \bibinfo{pages}{412--417}.
\newblock


\bibitem[Luglio et~al\mbox{.}(2007)]%
        {luglio_interworking_2007}
\bibfield{author}{\bibinfo{person}{Michele Luglio}, \bibinfo{person}{Cristiano Monti}, \bibinfo{person}{Cesare Roseti}, \bibinfo{person}{Antonio Saitto}, {and} \bibinfo{person}{Michael Segal}.} \bibinfo{year}{2007}\natexlab{}.
\newblock \showarticletitle{Interworking between {MANET} and satellite systems for emergency applications}.
\newblock \bibinfo{journal}{\emph{International Journal of Satellite Communications and Networking}} \bibinfo{volume}{25}, \bibinfo{number}{5} (\bibinfo{date}{Sept.} \bibinfo{year}{2007}), \bibinfo{pages}{551--558}.
\newblock
\showISSN{15420973, 15420981}
\href{https://doi.org/10.1002/sat.890}{doi:\nolinkurl{10.1002/sat.890}}


\bibitem[{Mattt}(2013)]%
        {mattt_multipeer_2013}
\bibfield{author}{\bibinfo{person}{{Mattt}}.} \bibinfo{year}{2013}\natexlab{}.
\newblock \bibinfo{title}{Multipeer {Connectivity}}.
\newblock
\urldef\tempurl%
\url{https://nshipster.com/multipeer-connectivity/}
\showURL{%
\tempurl}


\bibitem[Mesh(2024)]%
        {lochaMesh}
\bibfield{author}{\bibinfo{person}{Locha Mesh}.} \bibinfo{year}{2024}\natexlab{}.
\newblock \bibinfo{title}{Locha Mesh - The resilient network}.
\newblock
\urldef\tempurl%
\url{https://locha.io/}
\showURL{%
\tempurl}


\bibitem[Meshtastic(2024a)]%
        {meshtasticFirmware}
\bibfield{author}{\bibinfo{person}{Meshtastic}.} \bibinfo{year}{2024}\natexlab{a}.
\newblock \bibinfo{title}{Meshtastic device firmware}.
\newblock
\urldef\tempurl%
\url{https://github.com/meshtastic/firmware}
\showURL{%
\tempurl}


\bibitem[Meshtastic(2024b)]%
        {meshtasticMain}
\bibfield{author}{\bibinfo{person}{Meshtastic}.} \bibinfo{year}{2024}\natexlab{b}.
\newblock \bibinfo{title}{An open source, off-grid, decentralized, mesh network built to run on affordable, low-power devices}.
\newblock
\urldef\tempurl%
\url{https://meshtastic.org/}
\showURL{%
\tempurl}


\bibitem[Meshtastic(2024c)]%
        {meshtastic_RadioSettings}
\bibfield{author}{\bibinfo{person}{Meshtastic}.} \bibinfo{year}{2024}\natexlab{c}.
\newblock \bibinfo{title}{Radio Settings}.
\newblock
\urldef\tempurl%
\url{https://meshtastic.org/docs/overview/radio-settings/}
\showURL{%
\tempurl}


\bibitem[Meshtastic(2024d)]%
        {MeshtasticRTestConfig}
\bibfield{author}{\bibinfo{person}{Meshtastic}.} \bibinfo{year}{2024}\natexlab{d}.
\newblock \bibinfo{title}{Range Test Module Configuration}.
\newblock
\urldef\tempurl%
\url{https://meshtastic.org/docs/configuration/module/range-test/}
\showURL{%
\tempurl}


\bibitem[Preneel(2011)]%
        {preneel_cryptography_2011}
\bibfield{author}{\bibinfo{person}{Bart Preneel}.} \bibinfo{year}{2011}\natexlab{}.
\newblock \bibinfo{title}{Cryptography {Best} {Practices}}.
\newblock


\bibitem[Sciullo et~al\mbox{.}(2020)]%
        {LOCATE_Main}
\bibfield{author}{\bibinfo{person}{Luca Sciullo}, \bibinfo{person}{Angelo Trotta}, {and} \bibinfo{person}{Marco Di~Felice}.} \bibinfo{year}{2020}\natexlab{}.
\newblock \showarticletitle{Design and performance evaluation of a LoRa-based mobile emergency management system (LOCATE)}.
\newblock \bibinfo{journal}{\emph{Ad Hoc Networks}}  \bibinfo{volume}{96} (\bibinfo{year}{2020}), \bibinfo{pages}{101993}.
\newblock


\bibitem[Semtech(2024)]%
        {semtechWhatsLoraMain}
\bibfield{author}{\bibinfo{person}{Semtech}.} \bibinfo{year}{2024}\natexlab{}.
\newblock \bibinfo{title}{LoRa and LoRaWAN}.
\newblock
\urldef\tempurl%
\url{https://web.archive.org/web/20240328053451/https://lora-developers.semtech.com/documentation/tech-papers-and-guides/lora-and-lorawan/}
\showURL{%
\tempurl}


\bibitem[Shadbolt(2014)]%
        {shadbolt_firechat_2014}
\bibfield{author}{\bibinfo{person}{Peter Shadbolt}.} \bibinfo{year}{2014}\natexlab{}.
\newblock \bibinfo{title}{{FireChat} in {Hong} {Kong}: {How} an app tapped its way into the protests}.
\newblock
\urldef\tempurl%
\url{https://edition.cnn.com/2014/10/16/tech/mobile/tomorrow-transformed-firechat/index.html}
\showURL{%
\tempurl}


\bibitem[Sharevski et~al\mbox{.}(2020)]%
        {sharevski_beyond_2020}
\bibfield{author}{\bibinfo{person}{Filipo Sharevski}, \bibinfo{person}{Paige Treebridge}, \bibinfo{person}{Peter Jachim}, \bibinfo{person}{Audrey Li}, \bibinfo{person}{Adam Babin}, {and} \bibinfo{person}{Jessica Westbrook}.} \bibinfo{year}{2020}\natexlab{}.
\newblock \bibinfo{title}{Beyond {Trolling}: {Malware}-{Induced} {Misperception} {Attacks} on {Polarized} {Facebook} {Discourse}}.
\newblock
\urldef\tempurl%
\url{http://arxiv.org/abs/2002.03885}
\showURL{%
\tempurl}
\newblock
\shownote{arXiv:2002.03885 [cs]}.


\bibitem[Sharma et~al\mbox{.}(2023)]%
        {sharma2023lora}
\bibfield{author}{\bibinfo{person}{Deeksha~Rai Sharma}, \bibinfo{person}{Rohini~Ravindrasingh Raghuwanshi}, \bibinfo{person}{Tanvi Chandak}, {and} \bibinfo{person}{Dipali Ramdasi}.} \bibinfo{year}{2023}\natexlab{}.
\newblock \showarticletitle{LoRa-based IoT system for emergency assistance and safety in mountaineering.}
\newblock \bibinfo{journal}{\emph{International Journal of Safety \& Security Engineering}} \bibinfo{volume}{13}, \bibinfo{number}{3} (\bibinfo{year}{2023}).
\newblock


\bibitem[Un et~al\mbox{.}({[n.\,d.]})]%
        {ARPSPoC}
\bibfield{author}{\bibinfo{person}{Sok~Oeun Un}, \bibinfo{person}{Kimtho Po}, \bibinfo{person}{Kosorl Thourn}, \bibinfo{person}{Rathna Pec}, \bibinfo{person}{Channareth Srun}, {and} \bibinfo{person}{Seven Siren}.} \bibinfo{year}{[n.\,d.]}\natexlab{}.
\newblock \showarticletitle{Design of Emergency Position Reporting System for Disasters Using Amateur Radio and Automatic Packet Reporting System (APRS) as a Mobile Station Operator for Educational Purposes}.
\newblock \bibinfo{journal}{\emph{Indonesian Journal of Educational Research and Technology}} \bibinfo{volume}{3}, \bibinfo{number}{3} (\bibinfo{year}{[n.\,d.]}), \bibinfo{pages}{257--264}.
\newblock


\bibitem[Verma and Chauhan(2015a)]%
        {MANETCommunication}
\bibfield{author}{\bibinfo{person}{Himanshu Verma} {and} \bibinfo{person}{Naveen Chauhan}.} \bibinfo{year}{2015}\natexlab{a}.
\newblock \showarticletitle{MANET based emergency communication system for natural disasters}. In \bibinfo{booktitle}{\emph{International Conference on Computing, Communication \& Automation}}. IEEE, \bibinfo{pages}{480--485}.
\newblock


\bibitem[Verma and Chauhan(2015b)]%
        {verma_manet_2015}
\bibfield{author}{\bibinfo{person}{Himanshu Verma} {and} \bibinfo{person}{Naveen Chauhan}.} \bibinfo{year}{2015}\natexlab{b}.
\newblock \showarticletitle{{MANET} based emergency communication system for natural disasters}. In \bibinfo{booktitle}{\emph{International {Conference} on {Computing}, {Communication} \& {Automation}}}. \bibinfo{publisher}{IEEE}, \bibinfo{address}{Greater Noida, India}, \bibinfo{pages}{480--485}.
\newblock
\showISBNx{978-1-4799-8890-7}
\href{https://doi.org/10.1109/CCAA.2015.7148424}{doi:\nolinkurl{10.1109/CCAA.2015.7148424}}


\bibitem[ZH(2022)]%
        {cyrenMain}
\bibfield{author}{\bibinfo{person}{CYREN ZH}.} \bibinfo{year}{2022}\natexlab{}.
\newblock \bibinfo{title}{CYREN ZH: Cyber Resilience Network For The Canton Of Zurich}.
\newblock
\urldef\tempurl%
\url{https://dizh.ch/en/2022/07/07/cyren-zh-cyber-resilience-network-for-the-canton-of-zurich-2/}
\showURL{%
\tempurl}


\bibitem[Zimmerman and Ybarra(2016)]%
        {zimmerman_online_2016}
\bibfield{author}{\bibinfo{person}{Adam~G. Zimmerman} {and} \bibinfo{person}{Gabriel~J. Ybarra}.} \bibinfo{year}{2016}\natexlab{}.
\newblock \showarticletitle{Online aggression: {The} influences of anonymity and social modeling.}
\newblock \bibinfo{journal}{\emph{Psychology of Popular Media Culture}} \bibinfo{volume}{5}, \bibinfo{number}{2} (\bibinfo{date}{April} \bibinfo{year}{2016}), \bibinfo{pages}{181--193}.
\newblock
\showISSN{2160-4142, 2160-4134}
\href{https://doi.org/10.1037/ppm0000038}{doi:\nolinkurl{10.1037/ppm0000038}}


\end{thebibliography}

\appendix

\end{document}